\documentclass[10pt,journal,compsoc]{IEEEtran}
\IEEEoverridecommandlockouts

\usepackage[dvipsnames]{xcolor}
\usepackage{cite}
\usepackage{amsmath,amssymb,amsfonts}
\usepackage{algorithmic}
\usepackage{graphicx}
\usepackage{textcomp}
\usepackage{listings}
\usepackage{amsmath}
\usepackage{multirow}
\usepackage{enumitem}
\usepackage{hyperref}
\usepackage{url}
\usepackage{subcaption}

\def\BibTeX{{\rm B\kern-.05em{\sc i\kern-.025em b}\kern-.08em
		T\kern-.1667em\lower.7ex\hbox{E}\kern-.125emX}}

\usepackage{ifthen}
\usepackage{balance}
\newboolean{showcomments}
\setboolean{showcomments}{true}
\ifthenelse{\boolean{showcomments}}
{ \newcommand{\mynote}[2]{\textcolor{red}{
			\fbox{\bfseries\sffamily\scriptsize#1}
			{\small$\blacktriangleright$\textsf{\emph{#2}}$\blacktriangleleft$}}}}
{ \newcommand{\mynote}[2]{}}

\definecolor{diffstart}{named}{Blue}
\definecolor{diffincl}{named}{OliveGreen}
\definecolor{diffrem}{named}{Red}

\newcommand{\extrabold}{\bfseries}

\usepackage{listings}
\lstdefinelanguage{diff}{
	basicstyle=\ttfamily\extrabold\tiny,
	morecomment=[f][\color{diffstart}]{@},
	morecomment=[f][\color{diffincl}]{+},
	morecomment=[f][\color{diffrem}]{-},
	numbers=left,
	stepnumber=1,
	keepspaces=true,
	identifierstyle=\color{black},
}

\usepackage{pgfplots}
\pgfplotsset{width=3in,height=1.5in,compat=1.13}

\begin{document}
\pdfpagewidth 8.5in
\pdfpageheight 11.0in

\title{PatchNet: Hierarchical Deep Learning-Based Stable Patch Identification for the Linux Kernel}




\author{\IEEEauthorblockN{Thong Hoang\textsuperscript{1}, Julia Lawall\textsuperscript{2}, Yuan Tian\textsuperscript{3}, Richard J. Oentaryo\textsuperscript{4}, David Lo\textsuperscript{1}} \\
\IEEEauthorblockA{\textsuperscript{1}Singapore Management
  University/Singapore, \\ \textsuperscript{2}Sorbonne
  University/Inria/LIP6 \hspace{1cm}
\textsuperscript{3}Queen's University/Canada \\ \textsuperscript{4}McLaren Applied Technologies/Singapore} \\
vdthoang.2016@smu.edu.sg, Julia.Lawall@lip6.fr, yuan.tian@cs.queensu.ca, \\ richard.oentaryo@mclaren.com, davidlo@smu.edu.sg
}

\maketitle
\begin{abstract}

Linux kernel stable versions serve the needs of users who value stability
of the kernel over new features. The quality of such stable versions
depends on the initiative of kernel developers and maintainers to propagate bug fixing
patches to the stable versions. Thus, it is desirable to consider to what
extent this process can be automated. A previous approach
relies on words from commit messages and a small set of manually constructed code
features. This approach, however, shows only moderate accuracy.
In this paper, we investigate whether deep learning can provide a more
accurate solution. We propose PatchNet, a hierarchical deep learning-based approach capable of automatically extracting features from commit messages and commit code and using them to identify stable patches. PatchNet contains a deep hierarchical structure that mirrors the hierarchical and sequential structure of commit code, making it distinctive from the existing deep learning models on source code.
Experiments on 82,403 recent Linux patches confirm the superiority of   
PatchNet against various state-of-the-art baselines, including the one recently-adopted by Linux kernel maintainers. 
\end{abstract}

\section{Introduction}
\label{sec:intro}

The Linux kernel follows a two-tiered release model in which a
\textit{mainline} version, accepting bug fixes and feature enhancements, is
paralleled by a series of older {\em stable} versions that accept only bug
fixes~\cite{lee2003firm}.  The mainline serves the needs of users who want
to take advantage of the latest features, while the stable versions serve
the needs of users who value stability, or cannot upgrade their kernel due
to hardware and software dependencies. To ensure that there is as much
review as possible of the bug fixing patches and to ensure the highest
quality of the mainline itself, the Linux kernel requires that all patches
applied to the stable versions be submitted to and integrated in the
mainline first. A mainline developer or maintainer may identify a patch as
a bug fixing patch appropriate for stable kernels and add to the commit
message a Cc: stable tag (\textrm{stable@vger.kernel.org}). Stable-kernel
maintainers then extract such annotated commits from the mainline commit
history and apply the resulting patches to the stable versions that are
affected by the bug.


\textcolor{black}{A patch consists of a commit message followed by the code changes,
expressed as a unified diff~\cite{diffmanual:02}.  The diff consists of a series of
changes (removed and added lines of code), separated by lines beginning
with {\tt @@} indicating the number of the line in the affected source file
at which the subsequent change should be applied.  Each block of code
starting with an {\tt @@} line is referred to as a {\em hunk}.
Fig.~\ref{fig:sample_patch} shows three patches to the Linux kernel.  The
first patch changes various return values of the function {\tt
  csum\_tree\_block}.  The commit message is on lines 1-10 and the code
changes are on lines 11-25.  The code changes consist of multiple hunks,
only the first of which is shown in detail (lines 15-23).  In the shown
hunk, the function called just previously to the return site, {\tt
  map\_private\_extent\_buffer} (line 16), can return either 1 or a
negative value in case of an error.  So that the user can correctly
understand the reason for any failure, it is important to propagate such
return values up the call chain.  The patch thus changes the return value
of {\tt csum\_tree\_block} in this case from {\tt 1} to the value returned
by the {\tt map\_private\_extent\_buffer} call.  The remaining hunks
contain similar changes.  The Linux kernel documentation~\cite{stabledoc}
stipulates that a patch should be applied to stable kernels if it fixes a
real bug that can affect the user level and satisfies a number of criteria,
such as containing fewer than 100 lines of code and being obviously
correct.  This patch fits those criteria.  The patch was first included in
the Linux mainline version v4.6, and was additionally applied to the stable
version derived from the mainline release v4.5, first appearing in v4.5.5
(the fifth release based on Linux v4.5) as commit 342da5cefddb.}

\textcolor{black}{The remaining patches in Fig.~\ref{fig:sample_patch} should not be
propagated to stable kernels.  The patch in Fig.~\ref{fig:refactoring}
performs a refactoring, replacing some lines of code by a function call
that has the same behavior.  As the behavior is unchanged, there is no
impact on the user level.  The patch in Fig.~\ref{fig:nonstable} addresses
a minor performance bug, in that it removes some code that performs a
redundant operation.  The performance improvement should not be noticeable
at the user level, and thus this patch is not worth propagating to stable
kernels.  Note that none of the patches shown in
Fig.~\ref{fig:sample_patch} contains keywords such as ``bug'' or ``fix'',
or links to a bug tracking system.  Instead, the stable kernel maintainer
has to study the commit message and the code changes, to understand the
impact of the changes on the kernel code.}

\begin{figure}[t!]
\begin{subfigure}{\linewidth}
\begin{lstlisting}[language=diff]
commit 8bd98f0e6bf792e8fa7c3fed709321ad42ba8d2e
Author: Alex Lyakas <alex.bolshoy@gmail.com>
Date:   Thu Mar 10 13:09:46 2016 +0200

    btrfs: csum_tree_block: return proper errno value
    
    Signed-off-by: Alex Lyakas <alex@zadarastorage.com>
    Reviewed-by: Filipe Manana <fdmanana@suse.com>
    Signed-off-by: David Sterba <dsterba@suse.com>

diff --git a/fs/btrfs/disk-io.c b/fs/btrfs/disk-io.c
index d8d68af..87946c6 100644
--- a/fs/btrfs/disk-io.c
+++ b/fs/btrfs/disk-io.c
@@ -303,7 +303,7 @@ static int csum_tree_block(struct btrfs_fs_info *fs_info,
                err = map_private_extent_buffer(buf, offset, 32,
                                        &kaddr, &map_start, &map_len);
                if (err)
-                       return 1;
+                       return err;
                cur_len = min(len, map_len - (offset - map_start));
                crc = btrfs_csum_data(kaddr + offset - map_start,
                                      crc, cur_len);
@@ -313,7 +313,7 @@ static int csum_tree_block(struct btrfs_fs_info *fs_info,
...
\end{lstlisting}
\caption{A fix of a bug that can impact the user level.}
\label{fig:bugfix}
\end{subfigure}

\begin{subfigure}{\linewidth}
\begin{lstlisting}[language=diff]
commit 7b0692f1c60a9551f8ad5fe706b79a23720a196c
Author: Andy Shevchenko <...>
Date:   Wed Aug 14 11:07:11 2013 +0300

    HID: hid-sensor-hub: change kmalloc + memcpy by kmemdup
    
    The patch substitutes kmemdup for kmalloc followed by memcpy.
    
    Signed-off-by: Andy Shevchenko <...>
    Acked-by: Srinivas Pandruvada <...>
    Signed-off-by: Jiri Kosina <...>

diff --git a/drivers/hid/hid-sensor-hub.c b/drivers/hid/hid-sensor-hub.c
index 1877a2552483..e46e0134b0f9 100644
--- a/drivers/hid/hid-sensor-hub.c
+++ b/drivers/hid/hid-sensor-hub.c
@@ -430,11 +430,10 @@ static int sensor_hub_raw_event(struct hid_device *hdev,
			...
-                       pdata->pending.raw_data = kmalloc(sz, GFP_ATOMIC);
-                       if (pdata->pending.raw_data) {
-                               memcpy(pdata->pending.raw_data, ptr, sz);
+                       pdata->pending.raw_data = kmemdup(ptr, sz, GFP_ATOMIC);
+                       if (pdata->pending.raw_data)
                                pdata->pending.raw_size = sz;
-                       } else
+                       else
                                pdata->pending.raw_size = 0;
			...
\end{lstlisting}
\caption{A refactoring.}
\label{fig:refactoring}
\end{subfigure}

\begin{subfigure}{\linewidth}
\begin{lstlisting}[language=diff]
commit: 501bcbd1b233edc160d0c770c03747a1c4aa14e5 
Author: Thierry Reding <...>
Date:   Wed Apr 14 09:52:31 2014 +0200

    drm/tegra: dc - Do not touch power control register
    
    Setting the bits in this register is dependent on the output type driven
    by the display controller. All output drivers already set these properly
    so there is no need to do it here again.
    
    Signed-off-by: Thierry Reding <...>

diff --git a/drivers/gpu/drm/tegra/dc.c b/drivers/gpu/drm/tegra/dc.c
index 8b21e20..33e03a6 100644
--- a/drivers/gpu/drm/tegra/dc.c
+++ b/drivers/gpu/drm/tegra/dc.c
@@ -743,10 +743,6 @@ static void tegra_crtc_prepare(struct drm_crtc *crtc)
            WIN_A_OF_INT | WIN_B_OF_INT | WIN_C_OF_INT;
        tegra_dc_writel(dc, value, DC_CMD_INT_POLARITY);
-       value = PW0_ENABLE | PW1_ENABLE | PW2_ENABLE | PW3_ENABLE |
-           PW4_ENABLE | PM0_ENABLE | PM1_ENABLE;
-       tegra_dc_writel(dc, value, DC_CMD_DISPLAY_POWER_CONTROL);
        /* initialize timer */
        value = CURSOR_THRESHOLD(0) | WINDOW_A_THRESHOLD(0x20) |
            WINDOW_B_THRESHOLD(0x20) | WINDOW_C_THRESHOLD(0x20);
\end{lstlisting}
\caption{A fix of a minor performance bug.}
\label{fig:nonstable}
\end{subfigure}

\caption{Example patches to the Linux kernel.}
\label{fig:sample_patch}
\end{figure}

\textcolor{black}{As patches for stable kernels contain fixes for bugs that can impact the
user level, the quality of the stable kernels critically relies on the
effort that the developers and subsystem maintainers put into identifying
and labeling such patches, which we refer to as \textit{stable patches}.
This manual effort represents a potential weak point in the Linux kernel
development process, as the developers and maintainers may forget to label
some relevant patches, and apply different criteria for selecting them.
While the stable-kernel maintainers can themselves additionally pick up
relevant patches from the mainline commits, there are hundreds of mainline
commits per day, and many will likely slip past. This task can thus benefit
from automated assistance.}

\textcolor{black}{One way to provide such automated assistance is to build a tool that learns
from historical data how to differentiate stable from non-stable patches.
However, building such a tool poses some challenges. First, a patch contains
both a commit message (in natural language) and some code changes. While
the commit message is a sequence of words, and is thus amenable to existing
approaches on classifying text, the code changes have a more complex
structure.  Indeed, a single patch may include changes to multiple files;
the changes in each file consist of a number of hunks, and each hunk
contains zero or more removed and added code lines. As the structure of the
commit message and code changes differs, there is a need to extract their
features separately. Second, the historical information is noisy since
stable kernels do not receive only bug fixing patches, but also patches
adding new device identifiers and patches on which a subsequent bug fixing
patches depends. Moreover, patches that should have been propagated to
stable kernels may have been overlooked. Finally, as illustrated by
Fig.~\ref{fig:nonstable}, there are some patches that perform bug fixes but
should not be propagated to stable kernels for various reasons (e.g., lack
of impact on the user level or complexity of the patch).}

A first step in the direction of automatically identifying patches that
should be applied to stable Linux kernels was proposed by Tian et
al.~\cite{tian2012identifying} who combine LPU (Learning from Positive and
Unlabeled Examples)~\cite{letouzey2000learning} and SVM (Support Vector
Machine)~\cite{suykens1999least} to learn from historical information how
to identify bug-fixing patches.  Their approach relies on thousands of word
features extracted from commit messages and 52 features extracted from code
changes. The word features are obtained automatically by representing each
commit message as a bag of words, {\em i.e.}, a multiset of the words found
in the commit, whereas the code features are defined manually. The
bag-of-words representation of the commit message implies that the temporal
dependencies (ordering) of words in a commit message are ignored. The
manual creation of code features might overlook features that are important
to identify stable patches.

To address the limitations of the work of Tian et al.\ and to focus on
stable patches, we propose a novel hierarchical representation learning
architecture for patches, named PatchNet.  Like the LPU+SVM work, PatchNet
focuses on the commit message and code changes, as this information is
easily available and stable-kernel maintainers have reported to us that
they use one or both of these elements in assessing potential stable
patches. \textcolor{black}{Deviating from the previous LPU+SVM work, however, which requires
human effort to construct code features, PatchNet aims to automatically
learn two embedding vectors for representing the commit message and the set
of code changes in a given patch, respectively. While the first embedding
vector encodes the semantic information of the commit message to
differentiate between similar commit messages and dissimilar ones, the
latter embedding vector captures the sequential nature of the code changes
in the given patch.}  The two embedding vectors are then used to compute a
prediction score for a given patch, based on the similarity of the patch's
vector representation to the information learned from other stable or
non-stable patches. The key challenge is to accurately represent the
structure of code changes, which are not contiguous text like the commit
message, but rather amount to scattered fragments of removed and added code
across multiple files, within multiple hunks. Thus, different from existing
deep learning techniques working on source
code~\cite{white2016deep,huo2017enhancing,wang2016automatically,lam2017bug},
PatchNet constructs separate embedding vectors representing the removed
code and the added code in each hunk of each affected file in the given
patch. The information about a file's hunks are then concatenated to build
an embedding vector for the affected file. In turn, the embedding vectors
of all the affected files are used to build the representation of the
entire set of code changes in the given patch.

PatchNet has already attracted some industry attention. Inspired by the
work of Tian et al.\ and by our work on PatchNet, the Linux kernel stable
maintainer Sasha Levin has adopted a machine-learning based approach for
identifying patches for stable kernels, which we use as a baseline for our
evaluation (Section \ref{sec:baselines}). Recently Wen et
al.~\cite{wen2019ptracer} of ZTE Corporation have also adapted PatchNet to
the needs of their company.  These works show the potential usefulness of
PatchNet in an industrial setting.

The main contributions of this paper include:

\begin{itemize}[leftmargin=0.4cm]
\item We study the manual process of identifying patches for
  Linux stable versions. We explore the potential benefit of automatically
  identifying stable patches and summarize the challenges in using
  machine learning for this purpose.
 \item We propose a novel framework, PatchNet, to automatically learn a representation of a patch by considering both its commit message and corresponding code changes. PatchNet contains a novel deep learning model to construct an embedding vector for the code changes made by a patch, based on their sequential content and hierarchical structure. The two embedding vectors, representing the commit message and the set of code changes, are combined to predict whether a patch should be propagated to stable kernels.
\item We evaluate PatchNet on a new dataset that contains 82,403
  recent Linux
  patches. The results show the superiority of PatchNet compared to state-of-the-art baselines. PatchNet also achieves good performance on the complete set of Linux kernel patches. 
\end{itemize}

The rest of this paper is organized as
follows. Section~\ref{sec:background} introduces background
information. Section~\ref{sec:approach} elaborates the proposed PatchNet approach. Section~\ref{sec:exp} presents the experimental results. Section~\ref{sec:threat} discusses potential threats to validity. Section \ref{sec:related_work} highlights related work. Finally, Section~\ref{sec:conclusion} concludes and presents future work.

\section{Background}
\label{sec:background}
In this section, we present background information about the maintenance of
Linux kernel stable versions, the potential benefits of introducing
automation into the stable kernel maintenance process, and the challenges
posed for automation via machine learning.

\subsection{Context}
\label{sec:context}
\textcolor{black}{The Linux kernel, developed by Linus Torvalds in 1991, is a free and open-source, monolithic, and Unix-like operating system kernel~\cite{love2010linux}. It has been deployed on both traditional computer systems, i.e., personal computers and servers, and on many embedded devices such as routers, wireless access points, smart TVs, etc. Many devices, i.e., tablet computers, smartphones, smartwatches, etc.\ that have the Android operating system also use the Linux kernel.} 

\textcolor{black}{The Linux kernel includes a two tiered release model
  comprising a mainline version and a set of stable versions. The mainline
  version, often released every two to three months, is the version where
  all new features are introduced. After the mainline version is released, we consider it as ``stable''. Any bug fixing patches for a stable version are backported from the mainline version.}

Linux kernel development is carried out according to a hierarchical model,
with Linus Torvalds---who has ultimate authority about which
patches are accepted into the kernel---at the root and patch {\em authors} at the
leaves. A patch author is anyone who wishes to make a contribution to the
kernel, fix a bug, add a new functionality, or improve the coding
style. Authors submit their patches by email to {\em maintainers}, who
commit the changes to their git trees and submit pull requests up the
hierarchy. In this work, we are mostly concerned with the maintainers, who are responsible for 
assessing the correctness and usefulness of the
patches that they receive. Part of this responsibility involves determining
whether a patch is stable, and ensuring that it is annotated accordingly.

The Linux kernel provides a number of guidelines to help maintainers
determine whether a patch should be annotated for propagation to stable
kernels~\cite{stabledoc}. The main points are as follows:
\begin{itemize}[leftmargin=0.4cm]
\item It cannot be bigger than 100 lines.
\item It must fix a problem that causes a build error, an oops, a hang, data corruption,
a real security issue, or some ``oh, that’s not good'' issue.
\end{itemize}
These criteria may be simple, but are open
to interpretation. For example, even the criterion about patch size, which
seems unambiguous, is only satisfied by 93\% of the patches found in the
stable versions based on Linux v3.0 to v4.13, as of September
2017. 

\subsection{Potential Benefits of Automatically Identifying Stable Patches}
\label{seC:potential}

\begin{figure}
\begin{tikzpicture}
\begin{axis}[ybar,xtick=data,ticklabel style = {font=\scriptsize},
 label style = {font=\scriptsize},
 x tick label style={rotate=45,anchor=east, yshift=-0.01em},
ylabel={\begin{tabular}{c}\% commits\\propagated to stable\end{tabular}},
yticklabel=\pgfmathprintnumber{\tick}\,$\%$,
        ymin=0,
        ymax=30,
symbolic x
    coords={drivers/scsi,fs,arch/arm,mm,include,arch/x86,kernel,sound,drivers/usb,drivers/gpu,drivers/net,net}]
\addplot coordinates
{(drivers/scsi,                   8)
(fs,                             13)
(arch/arm,                       4)
(mm,                             13)
(include,                        4)
(arch/x86,                       10)
(kernel,                         11)
(sound,                          9)
(drivers/usb,                    17)
(drivers/gpu,                    7)
(drivers/net,                    6)
(net,                            11)};

\node[below,rotate=90,anchor=west] at (axis cs:drivers/scsi,8) {\scriptsize 519};
\node[below,rotate=90,anchor=west] at (axis cs:fs,13) {\scriptsize 531};
\node[below,rotate=90,anchor=west] at (axis cs:arch/arm,4) {\scriptsize 572};
\node[below,rotate=90,anchor=west] at (axis cs:mm,13) {\scriptsize 750};
\node[below,rotate=90,anchor=west] at (axis cs:include,4) {\scriptsize 804};
\node[below,rotate=90,anchor=west] at (axis cs:arch/x86,10) {\scriptsize 973};
\node[below,rotate=90,anchor=west] at (axis cs:kernel,11) {\scriptsize 1083};
\node[below,rotate=90,anchor=west] at (axis cs:sound,9) {\scriptsize 1323};
\node[below,rotate=90,anchor=west] at (axis cs:drivers/usb,17) {\scriptsize 1730};
\node[below,rotate=90,anchor=west] at (axis cs:drivers/gpu,7) {\scriptsize 2126};
\node[below,rotate=90,anchor=west] at (axis cs:drivers/net,6) {\scriptsize 2418};
\node[below,rotate=90,anchor=west] at (axis cs:net,11) {\scriptsize 2887};

\end{axis}
\end{tikzpicture}
\caption{Percentage of mainline commits propagated to stable kernels for the 12 directories with more than 500 mainline commits being propagated to stable kernels between Linux v3.0 (July 2011) and Linux v4.12 (July 2017). The number above each bar indicates the number of propagated commits.}
\label{pctprop}
\end{figure}
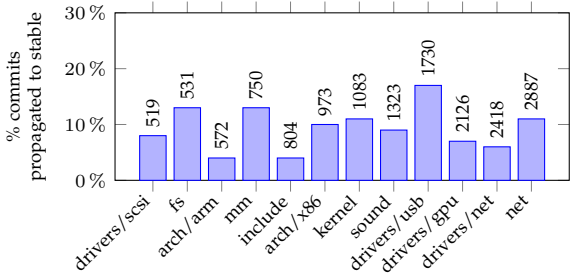

To understand the potential benefit of automatically identifying
stable patches, we examine the percentage of all mainline commits
that are propagated to stable kernels across different kernel subsystems
and the percentage of these that are annotated with the Cc: stable tag.  We
focus on the 12 directories for which more than 500 mainline commits were
propagated to stable kernels between Linux v3.0 (July 2011) and Linux v4.12
(July 2017).  Fig. \ref{pctprop} shows the percentage of all mainline
commits that are propagated to stable kernels for these 12 directories.  We
observe that there is a large variation in these values.  Comparing
directories with similar purposes, 4\% of {\tt arch/arm} (ARM hardware
support) commits are propagated, while 10\% of {\tt arch/x86} (x86 hardware
support) commits are propagated, and 6-8\% of the {\tt scsi}, {\tt gpu} and
{\tt net} driver commits are propagated, while 17\% of {\tt usb} driver
commits are propagated.\footnote{The {\tt usb} driver maintainer is also a
  stable kernel maintainer.} \textcolor{black}{If we make the assumption that the rate of bug introduction is
    roughly constant across similar kinds of code,
the wide variation in the propagation rates for similar kinds of code
    suggests that relevant commits may be being missed.}
    


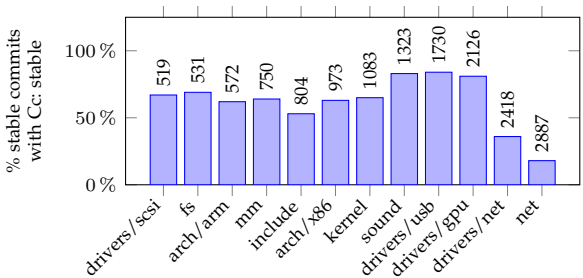
\begin{figure}
\begin{tikzpicture}
\begin{axis}[ybar,xtick=data,ticklabel style = {font=\scriptsize},
 label style = {font=\scriptsize},
 x tick label style={rotate=45,anchor=east, yshift=-0.01em},
ylabel={\begin{tabular}{c}\% stable commits\\with Cc: stable\end{tabular}},
yticklabel=\pgfmathprintnumber{\tick}\,$\%$,
        ymin=0,
        ymax=125,
symbolic x
    coords={drivers/scsi,fs,arch/arm,mm,include,arch/x86,kernel,sound,drivers/usb,drivers/gpu,drivers/net,net}]
\addplot coordinates
{(drivers/scsi,                   67)
(fs,                             69)
(arch/arm,                       62)
(mm,                             64)
(include,                        53)
(arch/x86,                       63)
(kernel,                         65)
(sound,                          83)
(drivers/usb,                    84)
(drivers/gpu,                    81)
(drivers/net,                    36)
(net,                            18)};

\node[below,rotate=90,anchor=west] at (axis cs:drivers/scsi,67) {\scriptsize 519};
\node[below,rotate=90,anchor=west] at (axis cs:fs,69) {\scriptsize 531};
\node[below,rotate=90,anchor=west] at (axis cs:arch/arm,62) {\scriptsize 572};
\node[below,rotate=90,anchor=west] at (axis cs:mm,64) {\scriptsize 750};
\node[below,rotate=90,anchor=west] at (axis cs:include,53) {\scriptsize 804};
\node[below,rotate=90,anchor=west] at (axis cs:arch/x86,63) {\scriptsize 973};
\node[below,rotate=90,anchor=west] at (axis cs:kernel,65) {\scriptsize 1083};
\node[below,rotate=90,anchor=west] at (axis cs:sound,83) {\scriptsize 1323};
\node[below,rotate=90,anchor=west] at (axis cs:drivers/usb,84) {\scriptsize 1730};
\node[below,rotate=90,anchor=west] at (axis cs:drivers/gpu,81) {\scriptsize 2126};
\node[below,rotate=90,anchor=west] at (axis cs:drivers/net,36) {\scriptsize 2418};
\node[below,rotate=90,anchor=west] at (axis cs:net,18) {\scriptsize 2887};

\end{axis}
\end{tikzpicture}
\caption{Percentage of mainline commits propagated to stable kernels that
  contain a Cc: stable tag for the 12 directories with more than 500 mainline commits being propagated to stable kernels between Linux v3.0 (July 2011) and Linux v4.12 (July 2017). 
  The number above each bar indicates the number of propagated commits.}
\label{pctcom}
\end{figure}

Fig. \ref{pctcom} shows the percentage of mainline commits propagated to
stable kernels that contain the Cc: stable tag, for the same set of kernel
directories.  The rate is very low for {\tt drivers/net} and {\tt net},
which are documented to have their own procedure \cite{stabledoc}.  The
others mostly range from 60\% to 85\%.  Commits in stable kernels that do
not contain the tag are commits that the stable kernel
maintainers have identified
on their own or that they have received via other non-standard channels.
This represents work that can be saved by an automatic labeling
approach.

\subsection{Challenges for Machine Learning}
\label{sec:challenges}
Stable patch identification poses some unique challenges for machine learning.
These include the kind of information available in a Linux kernel patch and the different reasons why patches are or are not selected for stable kernels.

First, patches contain a combination of text, represented by the commit
message, and code, represented by the enumeration of the changed
lines. Code is structured differently than text, and thus we need to
construct a representation that enables machine learning algorithms to
detect relevant properties. 

Second, the available labeled data from which to learn is somewhat
noisy. The only available source of labels is whether a given patch is
already in a stable kernel.  However, stable kernels in practice do not
receive only bug-fixing patches, but also patches that add new device
identifiers (structure field values that indicate some properties of a
supported device) and patches on which a subsequent bug-fixing patch
depends, as long as these patches are small and obviously correct.  On the
other hand, our results in the previous section suggest that not all
patches that should be propagated to stable kernels actually get propagated.  These
sources of noise may introduce apparent inconsistencies into the machine
learning process. 

\textcolor{black}{Finally, although some patches  perform bug fixes, not propagating them to stable kernels is the correct choice.}
One reason is that some parts of the code change so
rapidly that the patch does not apply cleanly to any stable
version. Another reason is that the bug was introduced since the most
recent mainline release, and thus does not appear in any stable version.

As the decision of whether to apply a patch to a stable kernel depends in part
on factors external to the patch itself, we cannot hope to achieve a perfect solution
based on applying machine learning to patches alone. 
Still, we believe that machine learning can effectively
complement existing practice by orienting stable-kernel maintainers towards
likely stable commits that they may have overlooked, even though
the above issues introduce the risk of some false negatives and false positives.

\subsection{Convolutional Neural Networks}
\label{sec:background_cnn}

\textcolor{black}{Convolutional Neural Networks (CNNs)~\cite{lecun1999object}
  are a class of deep learning models originally inspired by the
  connectivity pattern among neurons within an animal's visual cortex. Each cortical neuron responds to stimuli only in a restricted, local region of the visual field known as the receptive field. The receptive fields of different neurons partially overlap such that they cover the entire visual field. CNN has demonstrated successful applications in various problem domains, such as image/video recognition, natural language processing, recommender systems, etc.~\cite{karpathy2014large, lawrence1997face, krizhevsky2012imagenet}. A CNN typically has an input layer, a convolutional layer, followed by a nonlinear activation function, a pooling layer, a fully-connected layer, and an output layer. We briefly explain these layers in turn below.}

\textcolor{black}{The input layer often takes as an input a 2-dimensional
  matrix. The input is passed through a series of convolutional layers. The
  convolutional layers take advantage of the use of learnable
  filters. These filters are then applied along the entirety of the depth
  of the input data to produce a feature map.  The activation function is
  then applied to each value of the feature map. There are many types of
  activation function, i.e., sigmoid, hyperbolic tangent (tanh), rectified
  linear unit (ReLU), etc. In practice, most CNN architectures use ReLU as
  it achieves better performance compared to other activation
  functions~\cite{anastassiou2011univariate,dahl2013improving}.}

\textcolor{black}{After the application of the activation function, the pooling layer is 
    employed to reduce the dimensionality of the feature map and the number of parameters, which in turn helps mitigate data overfitting~\cite{tolias2015particular}. The pooling layer performs dimensionality reduction by applying an aggregation operation to the outputs of the feature map. There are three major types of pooling layer, i.e., max pooling layer, average pooling layer, and sum pooling layer, which use maximum, average and summation as aggregation operators, respectively. Among them, the max pooling is the most widely used in practice, as it typically demonstrates a better performance than the average or sum pooling~\cite{zeiler2013stochastic}.}
	
\textcolor{black}{The output of the pooling layer is often flattened and passed to a fully connected layer. The output of the fully connected layer then goes to an output layer, based upon which we can define a loss function to measure the quality of the outputs~\cite{lecun2015deep}. Accordingly, the goal of CNN training is to minimize this loss function, which is typically achieved using the stochastic gradient descent (SGD) algorithm~\cite{bottou2010large} or its variants.}

\section{Proposed Approach}
\label{sec:approach}

In this section, we first formulate the problem and provide an overview of PatchNet. We then describe the details of each module inside PatchNet. Finally, we present an algorithm for learning effective values of PatchNet's parameters.

\subsection{Framework Overview}
\label{sec:overall_framework}

The goal of PatchNet is to automatically label a
patch as stable or non-stable in order to reduce the manual effort for the stable-kernel maintainers. 
We consider the identification of stable patches as a learning task to construct a prediction function $f:
\mathcal{X} \longmapsto \mathcal{Y}$, where $\mathcal{Y} = \{0, 1\}$. Then, $x_i \in \mathcal{X}$ is identified as a stable patch when $f(x_i) = 1$.

\begin{figure}[!t]
	\center
	\includegraphics[scale=0.37]{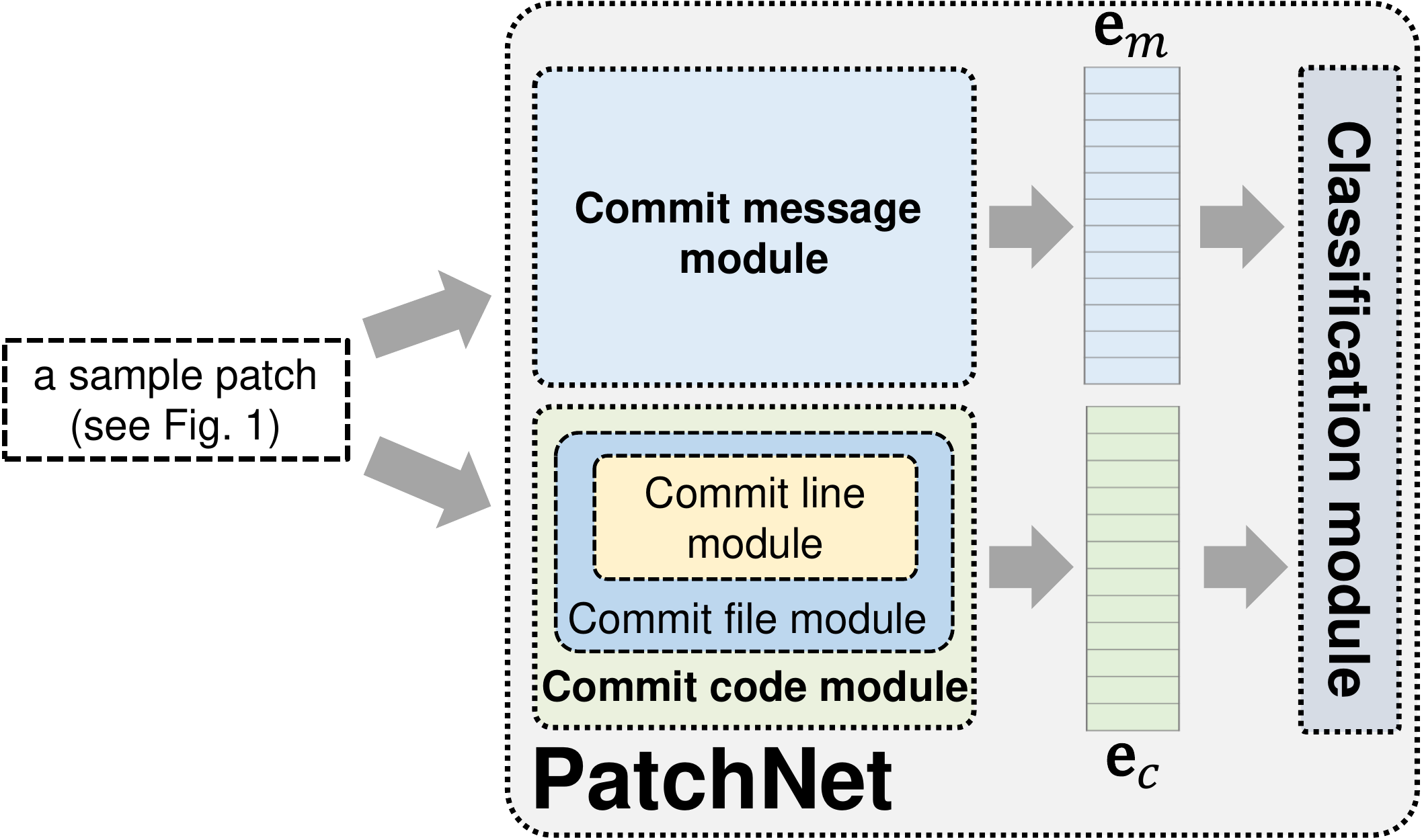}
	\caption{The proposed PatchNet framework. $\textbf{e}_m$ and $\textbf{e}_c$ are embedding vectors collected from the commit message module and commit code module, respectively.}
	\label{fig:patchnet}
\end{figure}

As illustrated in Fig.~\ref{fig:patchnet}, PatchNet consists of three
main modules: (1) a \textit{commit message module}, (2) a \textit{commit
  code module}, and (3) a \textit{classification module}.  The first two
are built upon a convolutional neural network (CNN)
architecture~\cite{lecun1998gradient, krizhevsky2012imagenet}, and aim
to learn a representation of the textual commit message
(cf. Fig.~\ref{fig:sample_patch}, lines 5-12) and the set of diff code
elements (cf. Fig.~\ref{fig:sample_patch}, lines 14-28) of a patch,
respectively.  The \textit{commit message module} and the \textit{commit
  code module} transform the commit message and the code changes into
embedding vectors $\mathbf{e}_m$ and $\mathbf{e}_c$, respectively. The two
vectors are then passed to the \textit{classification Module}, which computes a prediction score indicating the likelihood of a patch being a stable patch.

\subsection{Commit Message Module}
\label{sec:commit_msg_model}

\begin{figure}
\center
\includegraphics[scale=0.36]{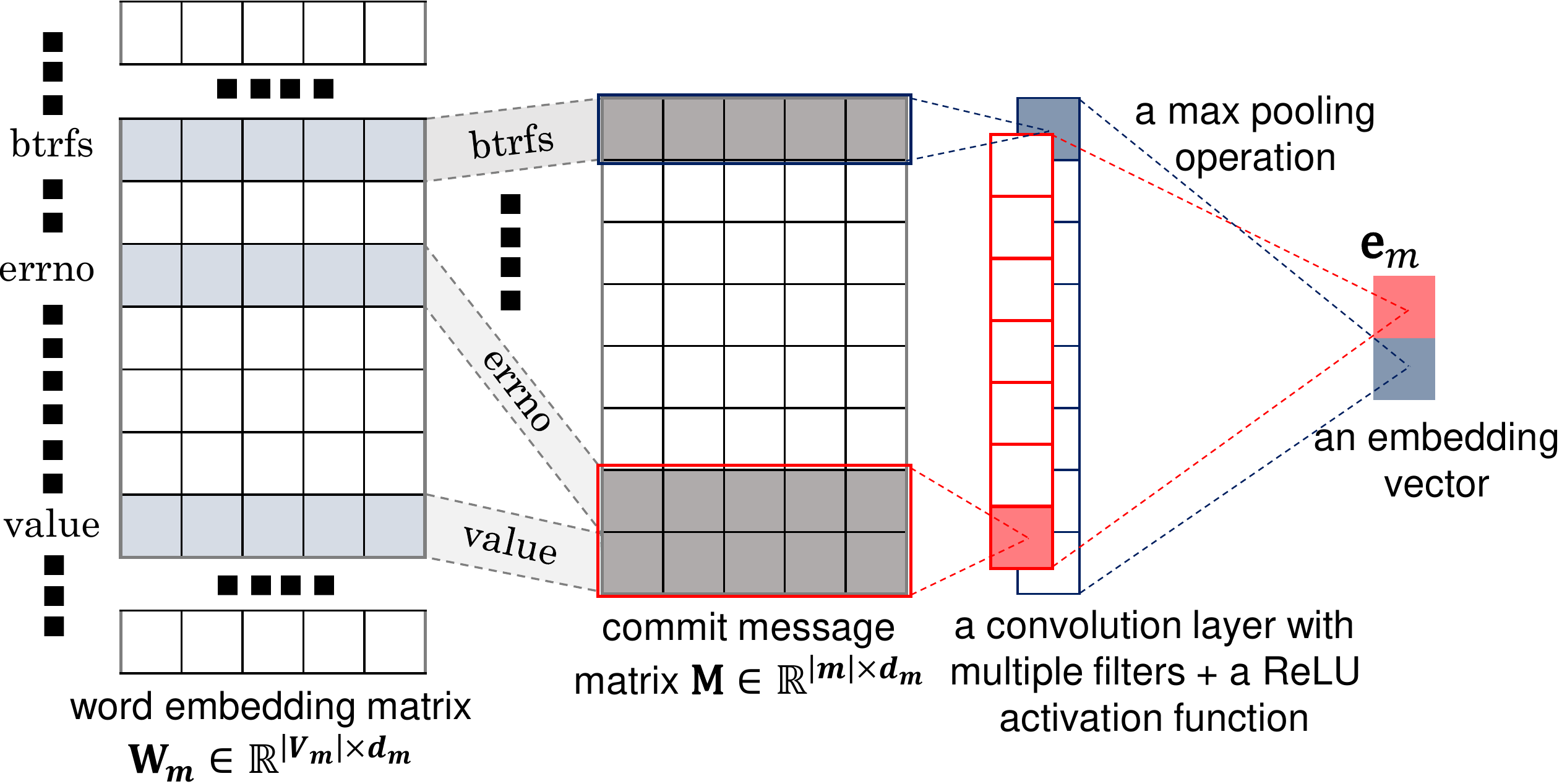}
\caption{Architecture of the commit message module. An embedding vector from the commit message is built by using a convolution layer with multiple filters and a max pooling operation.}
\label{fig:msg_model}
\end{figure}

Fig.~\ref{fig:msg_model} shows the architecture
of the commit message module, which is the same as the one proposed by Kim~\cite{kim2014convolutional} and Kalchbrenner \emph{\emph{\emph{et al.}}}~\cite{kalchbrenner2014convolutional} for sentence classification. 
The module involves an input message, represented as a two-dimensional matrix, a set of filters for identifying features in the message, and a means of combining the results of the filters into an \emph{embedding vector} that represents the most salient features of the message, to be used as a basis for classification.

\textbf{Message representation}. We encode a commit message as a
two-dimensional matrix by viewing the message as a sequence of vectors
where each vector represents one word appearing in the
message. The embedding vectors of the individual words are maintained using a
lookup table, the {\em word embedding matrix}, that is shared across all
messages.

Given a message $m$ as a sequence of words
$[\texttt{w}_1, \dots, \texttt{w}_{|m|}]$ and a word embedding matrix
$\textbf{W}_m \in \mathbb{R}^{|V_m|\times d_m}$, where $V_m$ is the vocabulary containing all words in commit messages and  $d_m$ is the
dimension of the representation of a word, the matrix representation
$\textbf{M} \in \mathbb{R}^{|m| \times d_m}$ of the message is:
\begin{equation} 
\textbf{M} =
[\textbf{W}[\texttt{w}_1], \dots, \textbf{W}[\texttt{w}_{|m|}]]
\end{equation}
For parallelization, all messages are padded or truncated to the same length.

\textbf{Convolutional layer}. The role of the convolutional layer is to apply
filters to the message, in order to identify the message's salient
features.  
A filter $\textbf{f} \in \mathbb{R}^{k \times d_m}$ is
a small matrix 
that is applied
to a window of $k$ words to produce a new feature. A feature $t_i$ is generated from a window of words
$\textbf{M}_{i:i+k-1}$ starting at word $i \leq |m|- k + 1$ by:
\begin{equation} 
\label{eq:filtering}
t_i = \alpha ( \textbf{f} \ast \textbf{M}_{i:i+k-1} + b_i) 
\end{equation}
where $\ast$ is a sum of element-wise products, $b_i \in
\mathbb{R}$ is a bias value, and $\alpha(\cdot)$ is a non-linear activation
function. For $\alpha(\cdot)$, we choose the {\em rectified linear unit} (ReLU) activation function~\cite{nair2010rectified}, as it has been shown to have better performance than 
its alternatives~\cite{anastassiou2011univariate,dahl2013improving}.
The filter $\textbf{f}$ is applied to all windows of size $k$ in the message resulting in a \textit{feature vector} $\textbf{t} \in \mathbb{R}^{|m|-k+1}$:
\begin{equation} 
\label{eq:ftr_vector}
\textbf{t} = [t_1, t_2, \cdots, t_{|m|-k+1}]
\end{equation}

\textbf{Max pooling}. 
To characterize the commit message, we are interested in the degree to
which it contains various features, but not where in the message those
features occur. Accordingly, for each filter,
we apply max pooling~\cite{collobert2011natural} over the feature vector $\textbf{t}$ to obtain the highest value:
\begin{equation} 
\label{eq:pooling}
\underset{1 \leq i \leq |m|-k+1}{\max} t_i
\end{equation}

\noindent
The results of applying max pooling to the feature vector resulting from
applying each filter are then concatenated 
unchanged to form an embedding vector
($\mathbf{e}_m$ on the right side of Fig.~\ref{fig:msg_model})
representing the meaning of the message.

\subsection{Commit Code Module}
\label{sec:commit_code_model}
Like the commit message,
the commit code can be viewed as a sequence of words. This view, however, overlooks the structure of code changes, as needed to distinguish between changes to different files, changes in
different hunks, and changes of different kinds (removals or additions). To
incorporate this structural information, PatchNet contains a {\em commit
  code module} that takes as input the code changes in a given patch and
outputs an embedding vector that represents the most salient features of
the code changes. The commit code module contains a {\em commit file
  module} that automatically builds an embedding vector representing the
code changes made to a given file in the patch. The embedding vectors of
code changes at file level are then concatenated unchanged into a single vector representing all the code changes made by the patch.

\subsubsection{Commit File Module}
\label{sec:code_file_framework}

The commit file module builds an embedding vector
for each file in the patch that represents the changes to the file.

\begin{figure}[t!]
	\center
	\includegraphics[scale=0.43]{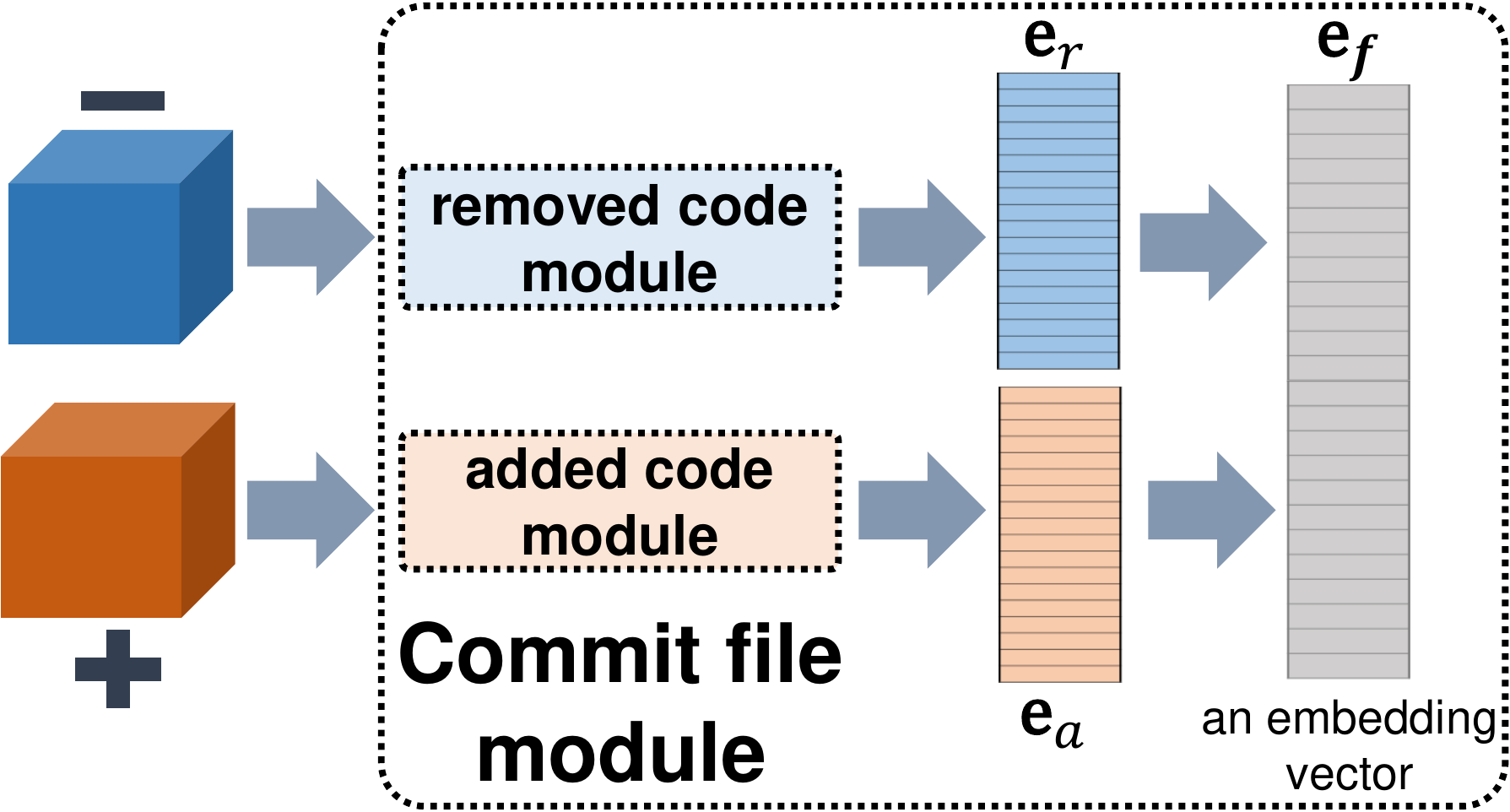}
	\caption{Architecture of the \textit{Commit File Module} for
          mapping a file in a given patch to an embedding vector. The input
          of the module is the removed code and added code of the affected
          file, denoted by ``--'' and ``+'', respectively.}
	\label{fig:commit_code_model}
\end{figure}

As shown in Fig.~\ref{fig:commit_code_model}, the commit file module
takes as input two matrices (denoted by ``--'' and ``+'' in
Fig.~\ref{fig:commit_code_model}) representing the removed code and added
code for the affected file in a patch, respectively. These two matrices
are passed to the \textit{removed code module} and the \textit{added code
module}, respectively, to construct corresponding embedding vectors. The
two embedding vectors are then concatenated unchanged to represent the code changes in each affected file. We present the removed code module and the added code module below.

\begin{figure*}[t!]
	\center
	\includegraphics[scale=0.38]{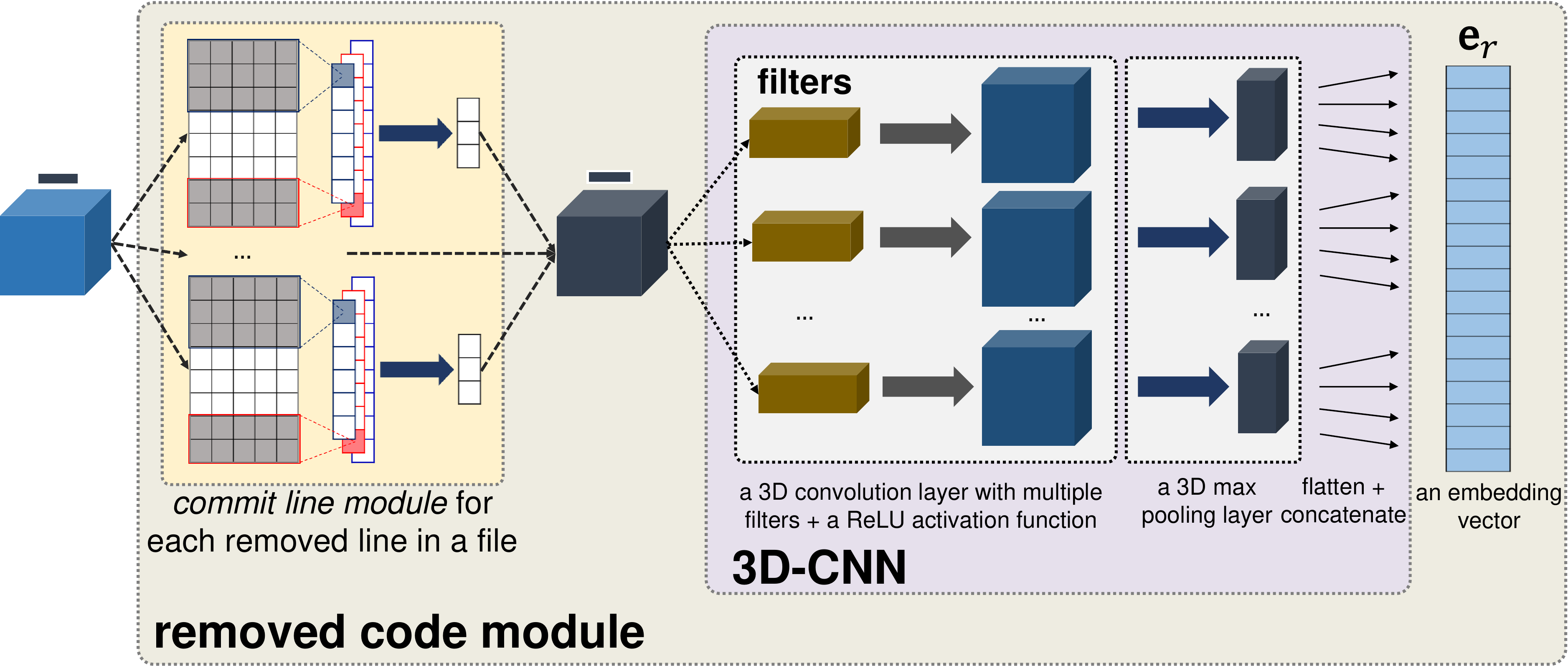}
	\caption{Architecture of the \textit{removed code module} used to build an embedding vector for the code removed from an affected file.}
	\label{fig:cnn3d}
\end{figure*}

\textbf{\textit{Removed code module.}} Fig.~\ref{fig:cnn3d}
shows the structure of the \textit{removed code module}. The
input of this module is a three-dimensional matrix, indicating the removed code
in a file of a given patch, denoted by $\mathcal{B}_r \in
\mathbb{R}^{\mathcal{H} \times \mathcal{N} \times \mathcal{L}}$, where
$\mathcal{H}$, $\mathcal{N}$, and $\mathcal{L}$ are the number of hunks,
the number of removed code lines for each hunk, and the number of words in each removed code line in the affected file, respectively. This module takes advantage of a {\em commit line module} and a 3D convolutional layer ({\em 3D-CNN}) to construct an embedding vector (denoted by $\textbf{e}_r$ in Fig.~\ref{fig:commit_code_model}) representing the removed code in the affected file. We describe the {\em commit line module} and the {\em 3D-CNN} in the following sections.

\textit{a) Commit line module.}  Each line of removed code in
$\mathcal{B}_r$ is processed by the {\em commit line module} to obtain a
list of embedding vectors representing the removed code lines. This module
has the same structure as the commit message module, but maintains a
code-specific vocabulary and word embedding matrix, as a word may have
different meanings in a textual message and in source code.

The obtained commit line vectors are used to construct a new three-dimensional matrix,
$\hat{\mathcal{B}}_r \in \mathbb{R}^{\mathcal{H} \times \mathcal{N} \times
  E}$. $\hat{\mathcal{B}}_r$ represents a sequence of ${\mathcal{H}}$
hunks; each hunk has a sequence of removed lines, where each line is now
represented as a $E$-dimensional embedding vector ($e_{ij} \in \mathbb{R}^E$) extracted by
the \textit{commit line module}. $\hat{\mathcal{B}}_r$ is then passed to the
3D convolutional neural network (3D-CNN), described below, to construct an embedding vector for the code removed from a file by a given patch.

\begin{figure}[t!]
	\center
    \includegraphics[scale=0.25]{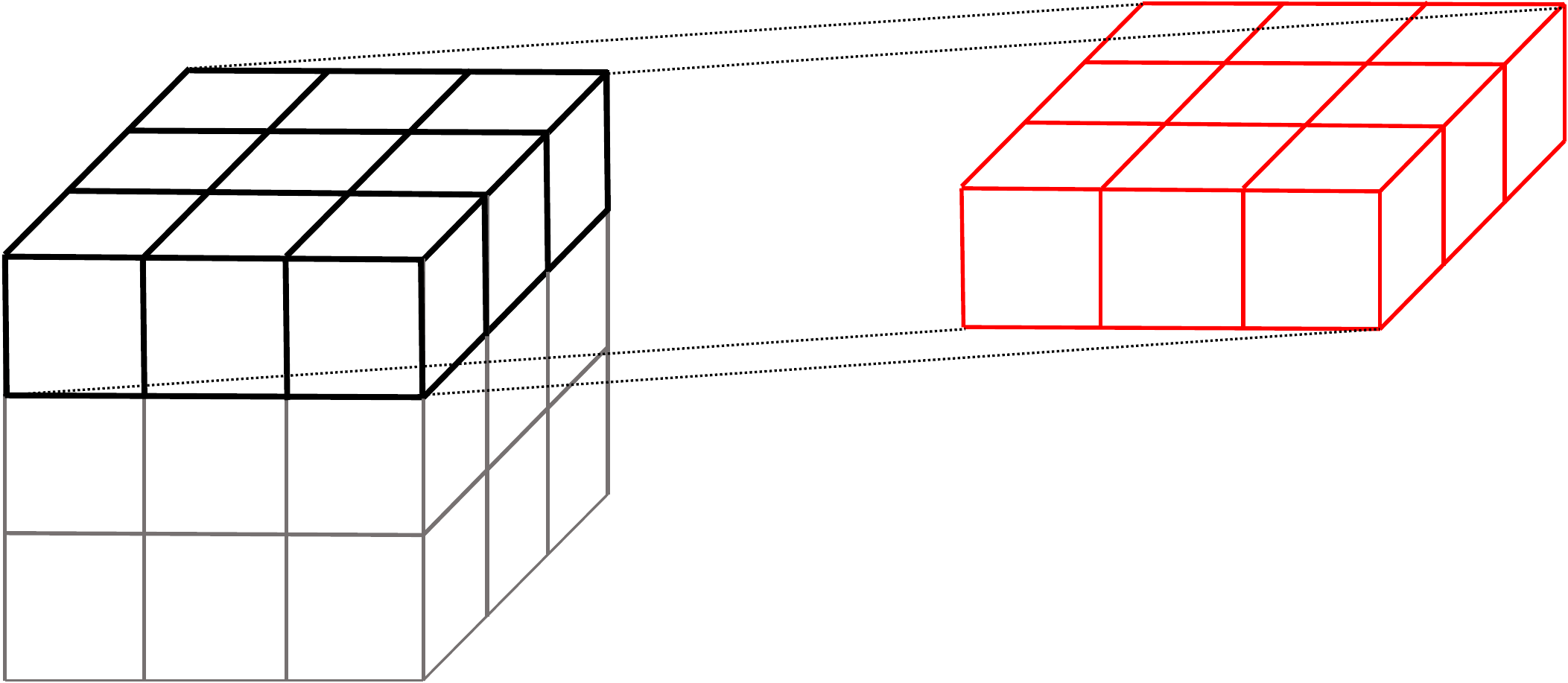}
	\caption{A 3D convolutional layer on $3 \times 3 \times 3$ data. The $1 \times 3 \times 3$ red cube on the right is the filter. The dotted lines indicate the sum of element-wise products over all three dimensions. The result is a scalar vector.}
	\label{fig:filtering_example}
   \vspace{-0.4cm}
\end{figure}

\textit{b) 3D-CNN}.
The 3D convolutional layer is used to
extract features from the code removed from the affected file, as
represented by $\hat{\mathcal{B}}_r$.  This layer applies each
filter $\textbf{F} \in \mathbb{R}^{k \times \mathcal{N} \times E}$
to a window of $k$ hunks $\textbf{H}_{i:i+k-1}$ to build a new feature as
follows:
\begin{equation} 
  \label{eq:3d_filter}
  f_i = \alpha(\textbf{F} \ast \textbf{H}_{i:i+k-1} + b_i)]
\end{equation}
$\ast$ is the sum of element-wise products, $\textbf{H}_{i:i+k-1} \in
\mathbb{R}^{|i:i+k-1| \times\mathcal{N} \times E}$ is constructed from the
$i$-th hunk through the $(i+k-1)$-th hunk in the removed code of the affected
file, $b_i \in \mathbb{R}$ is the bias value, and $\alpha(\cdot)$ is the
ReLU activation function. As for the commit message module (see Section~\ref{sec:commit_msg_model}), we
choose $k \in \{1, 2\}$. 
Fig.~\ref{fig:filtering_example} shows an example of a 3D convolutional layer that has one filter.
Applying the filter $\textbf{F}$ to all windows of hunks in
$\hat{\mathcal{B}_r}$ produces a feature vector: 
\begin{equation} 
\label{eq:ftr_hunk}
\mathcal{F} = [f_1, \dots, f_{\mathcal{H} - k + 1}]
\end{equation}

As in Section~\ref{sec:commit_msg_model}, we apply a max pooling operation
to $\mathcal{F}$ to obtain the most important feature.
The features selected by max pooling with multiple filters are concatenated
unchanged to
construct an embedding vector $\textbf{e}_r$ representing information
extracted from the removed code changes in the affected file.

\textbf{\textit{Added code module}}. This module has
the same architecture as the removed code module. The changes in the
added and removed code are furthermore padded or truncated to have the same
number of hunks ($\mathcal{H}$), number of lines for each hunk
($\mathcal{N}$), and the number of words of each line ($\mathcal{L}$), for
parallelization. Moreover, both modules also share the same vocabulary and
use the same word embedding matrix.

The added code module
constructs an embedding vector (denoted by $\textbf{e}_a$ in Fig.~\ref{fig:commit_code_model}) representing the added code in a file
of a given patch. An embedding vector representing all of the changes made to a given file by a commit is constructed by
concatenating the two embedding vectors representing the removed code and
added code as follows: 

\begin{equation} 
\label{eq:commit_file}
\textbf{e}_\texttt{f} = \textbf{e}_r \oplus \textbf{e}_a
\end{equation}  




\subsubsection{Embedding Vector for Commit Code}
\label{sec:construct_commit_code}
\textcolor{black}{The embedding vector for all the changes performed by a given patch is constructed as follows:}
\begin{equation} 
\label{eq:commit_code_vector}
\textbf{e}_c = \textbf{e}_{\texttt{f}_1} \oplus \dots \oplus \textbf{e}_{\texttt{f}_v}
\end{equation}
\textcolor{black}{where $\oplus$ simply concatenates two vectors unchanged, $\texttt{f}_i$ denotes the $i$-th file affected by the given commit, $v$ is the number of affected files, and $\textbf{e}_{\texttt{f}_i}$ denotes the vector constructed by applying the {\em commit file module} to the affected file $\texttt{f}_i$.}

\subsection{Classification Module}
\label{sec:classification_model}

\begin{figure}[t!]
	\center
	\includegraphics[scale=0.4]{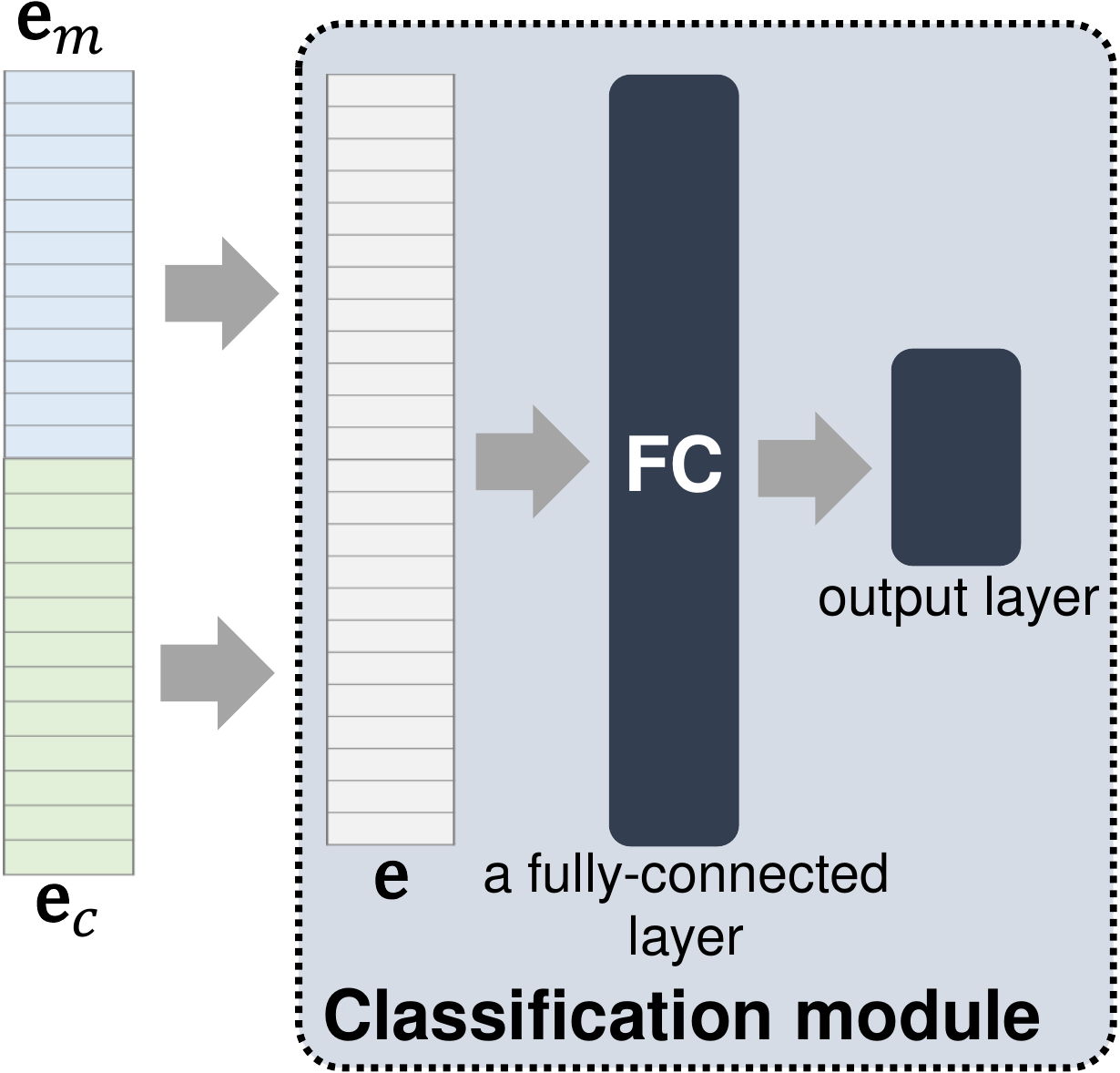}
	\caption{Architecture of the \textit{classification module}, comprising a fully connected layer (FC), and an output layer.}
	\label{fig:clf_module}
\end{figure}


Fig.~\ref{fig:clf_module} shows the architecture of the \textit{classification module}. It takes as input the commit message embedding vector $\mathbf{e}_m$ and the commit code embedding vector $\mathbf{e}_c$ discussed in Sections~\ref{sec:commit_msg_model} and~\ref{sec:commit_code_model}, respectively.
The patch is represented by their concatenation as follows: 

\begin{equation} 
\label{eq:commit_module}
\textbf{e} = \mathbf{e}_m \oplus \mathbf{e}_c
\end{equation}  

We then feed the concatenated vector $\mathbf{e}$ into a fully-connected
(FC) layer, which outputs a vector
$\mathbf{h}$ as follows:
\begin{equation}  
\label{eq:fully_layer}
\mathbf{h} = \alpha(\mathbf{w}_h \cdot \mathbf{e} + b_h)
\end{equation}
where $\cdot$ is a dot product, 
$\textbf{w}_h$ is a weight matrix associated with the concatenated vector,
$b_h$ is the bias value, and $\alpha(\cdot)$ is a non-linear
activation function. Again, we use ReLU to implement
$\alpha(\cdot)$. Note that both $\textbf{w}_h$ and $b_h$ are learned during our model's training process. Note that both $\textbf{w}_h$ and $b_h$ are learned during our model's training process.

Finally, the vector $\mathbf{h}$ is passed to an output layer, which
computes a probability score for a given patch:
\begin{equation}  
z_i = p(y_i=1|x_i) = \frac{1}{1 + \exp({-\mathbf{h} \cdot \mathbf{w}_o)}}
\end{equation}
where $\mathbf{w}_o$ is a weight matrix 
that is also learned during the training process. 


\subsection{Parameter Learning}
\label{sec:learning}

During the training process, PatchNet learns the following parameters: the
word embedding matrices for commit messages and commit code, the filter matrices and bias of the convolutional layers, and the weights and bias of
the fully connected layer and the output layer. The training aims to minimize the following 
regularized loss function~\cite{haykin2001kalman}:
\begin{equation} 
\label{eq:cost}
\begin{split}
\mathcal{O} &= -\log\left( \prod_{i=1}^{N} p(y_i|x_i) \right) + \frac{\lambda}{2} \|\theta\|_{2}^{2} \\
 &= -\sum_{i=1}^{N} \left[ y_i \log (z_i) + (1 - y_i) \log(1 - z_i) \right] + \frac{\lambda}{2} \|\theta\|_{2}^{2}
\end{split}
\end{equation}
where $z_i$ is the probability score from the output layer and $\theta$
contains all the (learnable) parameters as mentioned before. 

The term $\frac{\lambda}{2} \|\theta\|_{2}^{2}$ is used to mitigate data overfitting by penalizing large model parameters, thus reducing the model complexity. 
To further improve the robustness of our model, we also apply the dropout technique~\cite{srivastava2014dropout} on all the convolutional and fully-connected layers in PatchNet. 


To minimize the regularized loss function (\ref{eq:cost}), we employ a
variant of stochastic gradient descent (SGD)~\cite{bottou2010large}
called \textit{adaptive moment estimation}
(Adam)~\cite{kingma2014adam}. We choose Adam over SGD due to its computational efficiency and low memory requirements~\cite{kingma2014adam, anthimopoulos2016lung, arora2018optimization}.  

\section{Experiments}
\label{sec:exp}

We first describe our dataset and how we preprocess it. We
then introduce the baselines and evaluation metrics. Finally, we present
our research questions and results.

\subsection{Dataset}
\label{sec:data}

We take our data from the patches that have been committed to mainline
Linux
kernel\footnote{git://git.kernel.org/pub/scm/linux/kernel/git/torvalds/linux.git}
v3.0, released in July 2011, through v4.12, released in July 2017.  We
additionally collect information from the stable
kernels\footnote{git://git.kernel.org/pub/scm/linux/kernel/git/stable/linux.git}
that had been released as of October 2017 building on Linux kernels v3.0
through v4.13. \textcolor{black}{We consider a mainline commit to be stable if it is
  duplicated in at least one stable version.  To increase the set of
  commits that can be used for training, we furthermore include in the
  training set of stable patches other Linux kernel commits that are
  expected by convention to be bug-fixing patches.  Indeed, a Linux kernel
  release is created by first collecting a set of commits for the coming
  release into a preliminary release called a ``release candidate'', named
  {\em rc1}, that may include new features and bug fixes.  This is followed
  by a succession of further release candidates, named {\em rc2} onwards,
  that should include only bug fixes.  We thus also include the commits
  added for release candidates rc2 onwards in our set of stable patches.}

We refer to patches that are propagated to stable kernels or
are found in later release candidates as {\em stable patches} and patches
that are not propagated to stable kernels or found in later release
candidates as {\em non-stable patches}.  To avoid biasing the learning
process towards either stable or non stable patches, we construct our training
datasets such that the number of patches in each category is roughly
balanced.  While this situation does not reflect the number of stable and
non-stable patches that confront a stable kernel maintainer each day, it
allows effective training and interpretation of the experimental results.

\subsubsection{Identifying Stable Patches}
\label{sec:stable_patch}

The main challenge in constructing the datasets is to determine which
mainline patches have been propagated to stable kernels.  Indeed, there is
no required link information connecting the two. Many stable patches
explicitly mention the corresponding mainline commit in the commit message,
which we refer to as a {\em back link}. For others, we rely on the author
name and the subject line. Subject lines typically contain information
about both the change made and the name of the file or directory in which
the change is made, and should be unique.  We first
collect from the patches in the stable kernels a list of back
links and a list of pairs of author name and subject line.  A commit from
the mainline whose commit id is mentioned in a back link or whose author
name and subject line are the same as one found in a patch to a stable
kernel is considered to be a stable patch.

\subsubsection{Collecting the Dataset}
\label{sec:mainline_patch}

We collect our dataset from the mainline Linux kernel.  In order to focus on
patches that are challenging for stable maintainers to classify, we drop in
advance all patches that do not meet the stable-kernel size
guidelines,\footnote{https://www.kernel.org/doc/html/v4.15/process/stable-kernel-rules.html}
{\em i.e.}, those that exceed 100 code lines, including both changed lines
and context as reported by {\tt diff}.  We subsequently keep all
identified stable patches for our dataset and select an equal number of non-stable patches.  Whenever possible, we select non-stable patches that have a
similar number of changed lines as the stable patches, again to create a
dataset that reflects the cases that cannot be excluded by size alone and
thus are challenging for stable kernel maintainers.  These patches are then
subject to a preprocessing step that is detailed in the next
section. We do not use the dataset studied by Tian \emph{et al.}~\cite{tian2012identifying}, because it is seven years old and unclean, including labeling as bug-fixing patches the results of tools that may report coding style issues or faults whose impact is not visible in practice.

Our dataset comes from Linux kernel mainline versions 3.0 (July 2011)
through 4.12 (July 2017). There were 424,380 commits during that period.  We
consider only those commits that are not merge commits, that modify a file
as opposed to only adding or removing files, and that affect at least one
{\tt .c} or {\tt .h} file.  This leaves 346,570 commits (82\%).  Of these
346,570 commits, 79,319 (23\%) are not considered because they contain more
than 100 changed lines, leaving 267,251 commits. Of these,to have a
balanced training dataset, we pick the
42,408 stable patches for which the preprocessing step is successful (see
below,) and
39,995 non-stable patches, {\em i.e.}, 82,403 patches in all. In RQ4, we
consider the full set of Linux kernel patches in versions v3.0-v4.12 that
are accepted by our preprocessing step.


\subsection{Patch Preprocessing}
\label{sec:patch_processing}

Our approach applies some preprocessing steps to the patches before they are
given to PatchNet.

\subsubsection{Preprocessing of Commit Messages}
\label{sec:preprocessing_msg}

Our approach applies various standard natural language techniques to the
commit messages, such as stop word elimination and
stemming~\cite{vijayarani2015preprocessing, brants2003natural}, to reduce
message length and eliminate irrelevant information.  Subsequently, we pad
or truncate all commit messages to the same size, specifically 512 words,
covering the complete commit message for all 
patches, for
parallelism.  Because we are interested in cases that are challenging for
the stable kernel maintainer, we drop tags such as Cc: stable and Fixes,
whose goal is to indicate that a given patch is a stable or a bug
fixing patch.  We also drop tags indicating who has approved the patch, as
the set of developers and their work profiles can change in the future.

\subsubsection{Preprocessing of Code Changes}
\label{sec:extract_level_diff}

Diff code elements, as illustrated in Fig.~\ref{fig:bugfix}, may have
many shapes and sizes, from a single word to multiple lines spread out over
multiple hunks. To describe changes in terms of meaningful syntactic units
and to provide context for very small changes, we collect differences at
the granularity of atomic statements. These may be, {\em e.g.}, simple
assignment statements, return statements, if headers, etc.  For example, in
the patch illustrated in Fig.~\ref{fig:bugfix}, the only change is to
replace {\tt 1} on line 22 by {\tt err} on line 23.  Nevertheless, we
represent the change as a change in the complete return statement, {\em
  i.e.}, {\tt return 1;} that is transformed into {\tt return err;}. \textcolor{black}{We
also distinguish changes in error checking code (code to detect whether an
error has occurred, {\em e.g.}, line 21 in Fig.~\ref{fig:bugfix}) and
in error handling code (code to clean up after an error has occurred, {\em
  e.g.}, lines 22 and 23 in Fig.~\ref{fig:bugfix}) from changes in
other code, which we refer to as {\em normal code}. Error handling code is
considered to be any code that is in a conditional with only one branch,
where the conditional ends in a {\tt return} with an argument other than
{\tt 0} (0 is typically the success indicator) or a {\tt goto}, as well as any code following a label that ends in
a {\tt return} with an argument other than {\tt 0} or a {\tt goto}.  Error
checking code is considered to be the header of a conditional that matches
the former pattern.  These criteria are not completely reliable, as such
code can sometimes represent the success case rather than a failure case,
but they are typically followed and are actively promoted by Linux kernel
developers.  Error checking code and error handling code are very common in
the Linux kernel, which must be robust, and they are disjoint in structure
and purpose from the implementation of the main functionality.}

For a given commit, the first step is to extract the names of the affected
files and to extract the state of those files before and after the
commit. Analogous to the stemming and stop word elimination performed at
the commit message level, for each before and after file instance, we
remove comments and the contents of strings, as changes in comments and
within strings are not likely to be needed in stable kernels. For a given pair of
before and after files, we then compute the difference using the command
``git diff -U0 old new'', giving the changed lines with no lines of
surrounding context. For each ``--'' or ``+'' line in the diff output, we
then collect a record indicating the sign (``--'' or ``+''), the category
(error-handling code, etc.), the hunk number, the line number in the old or
new version, respectively, and the starting and ending columns of the
non-space changes on the line.  We furthermore keep the names of called
functions, when these are not defined in the same file and are used at
least 5 times, but drop other identifiers, {\em i.e.} field names and
variable names, as these may be too diverse to allow effective learning and unnecessarily slow down the training time.
Indeed, adding just the frequently used function names increases the code
vocabulary size from 43 to 3,616 unique tokens, which
increases the training time.

To extract changes at the level of atomic statements, rather than the
individual lines obtained by diff, we parse each file as it exists before
and after the change and keep the atomic statements that intersect with
a changed line observed by diff.  For this, we use the parser of the C
program transformation system Coccinelle~\cite{padioleau2008documenting},
which uses heuristics to parse around compiler directives and macros
\cite{yoann:cc}.  This makes it possible to reason about patches in terms
of the way they appear to the user, without macro expansion, but comes with
some cost, as some patches must be discarded because the parsing heuristics
are not sufficient to parse all of the code affected by the changed
lines. 

By following the above-mentioned steps, we collect the files affected by a
given patch. For each removed or added code line of an affected file,
denoted by ``--'' and ``+'', we collect the corresponding hunk number and line number. Each word in a line is a
pair of the associated token and the annotation indicating whether the word
occurs on a line of as error-checking code, error-handling code, or normal
code.  This information is used to build the two three-dimensional matrices
representing the removed code and the added code for the affected file
(see Fig. \ref{fig:commit_code_model}).

\subsection{Baselines}
\label{sec:baselines}
We compare PatchNet with several baselines:

\begin{itemize}[leftmargin=0.4cm]
\item \emph{Keyword}: As a simple but frequently used
  heuristic~\cite{tian2012identifying}, we select all commits in which the
  commit message includes ``bug'', ``fix'', or ``bug-fix'' after conversion
  of all words to lowercase and stemming.  While not all bug fixes are
  relevant for stable kernels, as some bugs may have very low impact or the
  fix may be too large or complex to be considered clearly correct, the
  problem of identifying bug fixes is close enough to that of recognizing
  stable patches to make comparison with our model valuable.
\item \emph{LPU+SVM}: This method was proposed
  by Tian et
  al.~\cite{tian2012identifying} and combine
  Learning from Positive and Unlabeled Examples
  (LPU)~\cite{joachims1999svmlight, letouzey2000learning, liu2003building}
  and Support Vector Machine (SVM)~\cite{cauwenberghs2001incremental,
    cristianini2000introduction}, to build a classification
  model for automatically identifying bug fixing patches. The set of code
  features considered was manually selected. In Tian \emph{et al.}'s work, stable kernels were
  considered as a source of bug-fixing patches in the training and testing
  data.

 \item \emph{LS-CNN}: 
    \textcolor{black}{Huo \emph{et al.}~\cite{huo2017enhancing} combined
      LSTM~\cite{hochreiter1997long} and CNN~\cite{lecun1998gradient} to
      localize potential buggy source files based on bug report
      information. They used CNN to learn a representation of the bug
      report and a combination of LSTM and CNN to learn the structure of
      the code. To assess the ability of LS-CNN to classify patches as
      stable, for a given patch, we give the commit message and the code
      changes (i.e., the result of concatenating the lines changed in the
      various files and hunks) as input to LS-CNN in place of the bug
      report and the potential buggy source file, respectively. To make a
      fair comparison, the CNN used to learn the representation of the
      commit message in LS-CNN has the same architecture (i.e., number of
      convolutional layer, filter size, activation function, etc.) as the
      CNN used to learn the representation of the commit message in PatchNet.}

\item \emph{Feed-forward fully connected neural network} (F-NN): Inspired by PatchNet
  and the work of Tian \emph{et al.}\ on LPU+SVM, a Linux stable kernel
  maintainer, Sasha Levin, has developed an approach to identifying
  stable patches \cite{sasha} based on a feed-forward fully
  connected neural network~\cite{bishop1995neural, goodfellow2016deep} and
  a set of manually selected features, including frequent commit message
  words, author names, and some code metrics.  Levin actively uses this
  approach in his work on the Linux kernel.
\end{itemize}

For LPU-SVM and LS-CNN, we used the same parameters and settings as
described in the respective papers. For F-NN, we asked Levin to train the tool on our training data and
test it with our testing data. We use 50\% as the cut off for considering a patch as stable for PatchNet and all baselines.

\subsection{Experimental Settings}
\label{sec:parameters}

  
\textcolor{black}{PatchNet has several hyperparameters (i.e., the sizes of the filters, the number of convolutional filters, the size of the fully-connected layer, etc.) that we instantiate them in the following paragraph.}

For the sizes of the filters described in Section~\ref{sec:approach}, we choose $k \in \{1, 2\}$, making the associated
windows analogous to a 1-gram or 2-gram as used in natural language
processing~\cite{jurafsky2014speech, brown1992class}. 
  Using 2-grams
allows our approach to take into account the temporal ordering of words,
going beyond the bag of words used by Tian et
al.~\cite{tian2012identifying}. The number of convolutional filters is set
to 64. The size of the fully-connected layer described in Section~\ref{sec:classification_model} is set to 100. The
dimensions of the word vectors in commit message $d_m$ and code changes
$d_c$ are set to 50. PatchNet is trained using Adam~\cite{kingma2014adam}
with shuffled
mini-batches. The batch size is set to 32. We train PatchNet for 50 epochs
and apply the early stopping strategy~\cite{prechelt1998automatic, caruana2001overfitting}, i.e.,
we stop the training if there has been no update to the loss value (see
Equation~\ref{eq:cost}) for the last 5 epochs. All these hyperparameter values
are widely used in the deep learning community~\cite{severyn2015learning, huo2016learning, huo2017enhancing, hinton2012improving}.  For parallelization,  the  number  of  changed  files,  the  number  of hunks  for  each  file,  the  number  of  lines  for  each  hunk,  the number of words of each removed or added code are set to 5, 8, 10, and 120, respectively.

\textcolor{black}{In our experiments, we run PatchNet on Ubuntu 18.04.3 LTS, 64 bit, with a Tesla P100-SXM2-16GB5 GPU.\footnote{https://www.nvidia.com/en-us/data-center/tesla-p100/} Training takes around 20 hours and testing less than 30 minutes to process 16,481 patches (one of the five folds presented in Section~\ref{sec:rq}). Note that training only needs to be done periodically (e.g., weekly/monthly) and the trained model can be used to label many patches. In our experiments, on average, the trained PatchNet can assign a label to a single patch in 0.11 seconds.}




\subsection{Evaluation Metrics}
\label{sec:metrics}
To evaluate the effectiveness of a stable patch identification
model, we employ the following metrics:

\begin{itemize}[leftmargin=0.4cm]
\item \emph{Accuracy}: Proportion of stable and non-stable patches that are correctly classified.
\item \emph{Precision}: Proportion of patches that are correctly classified as stable. 
\item \emph{Recall}: Proportion of stable patches that are correctly classified. 
\item \emph{F1 score}: Harmonic mean between precision and recall
\item \emph{AUC}: Area under the Receiver Operating Characteristic curve,  measuring if the stable patches tend to have higher predicted probabilities (to be stable) than non-stable ones. 
\end{itemize}


\subsection{Research Questions and Results}
\label{sec:rq}
Our study seeks to answer several research questions (RQs): 


{\bf RQ1: Do the properties of stable and non-stable patches change over time?}
A common strategy for evaluating machine learning
algorithms is $n$-fold cross-validation~\cite{kohavi1995study},
in which a dataset is randomly distributed among $n$ equal-sized buckets,
each of
which is considered as test data for a model trained on the
remaining $n-1$ buckets.  When data elements become available over time, as
is the case of Linux kernel patches, this strategy results in testing a
model on data that predates some of the data on which the model was
trained.  Respecting the order of patch submission,
however, would limit the amount of testing that can be done, given
the fairly small number of stable patches available.

To address this issue, we first assess whether training on future data
helps or harms the accuracy of PatchNet.  We first sort the
patches collected in Section~\ref{sec:data} from earliest to latest based
on the date when the patch author submitted the patch to maintainers. Then,
we divide the dataset into five mutually exclusive sets by date. Note that the
resulting five sets are not perfectly balanced, but they come close, with
stable patches making up 45\% to 55\% of each set.  Then, we repeat the
following process five times: take one set as a testing set and use the
remaining four sets for training. Testing on the first set shows
the impact of training only on future data.  Testing on the fifth set
shows the impact of training only on past data.  The other testing
sets use models trained on a mixture of past and future data.

Table~\ref{tab:cross_valid_patchnet} shows the results of PatchNet on the
different test sets. The standard deviations are quite small (i.e., at most
0.013), hence there is no difference between training on past or future
data.  Our dataset starts with Linux v3.0, which was released in 2011,
twenty years after the start of work on the Linux kernel.  The lack of
impact due to training on past or future data suggests that in such a
mature code base the properties that make a patch relevant for stable
kernels are fairly constant over time.  This property is indeed beneficial,
because it means that our approach can be used to identify stable commits that have been missed in older versions.  In the subsequent research
questions, we thus retain the same five test and training sets.

\begin{table}[t!]
  \centering
  \caption{The results of PatchNet on the five chronological
    test sets}
    \begin{tabular}{|l|c|c|c|c|c|}
    \hline
          & \textbf{Accuracy} & \textbf{Precision} & \textbf{Recall} & \textbf{F1}    & \textbf{AUC} \\
    \hline
    \hline
    Set=1 & 0.852 & 0.841 & 0.886 & 0.863 & 0.850 \\
    \hline
    Set=2 & 0.860  & 0.833 & 0.909 & 0.869 & 0.859 \\
    \hline
    Set=3 & 0.866 & 0.833 & 0.910  & 0.870  & 0.867 \\
    \hline
    Set=4 & 0.864 & 0.828 & 0.912 & 0.868 & 0.864 \\
    \hline
    Set=5 & 0.869 & 0.860  & 0.917 & 0.887 & 0.862 \\
    \hline
    \hline
    \textbf{Std.} & \textbf{0.007} & \textbf{0.013} & \textbf{0.012} & \textbf{0.009} & \textbf{0.007} \\
    \hline
    \end{tabular}%
  \label{tab:cross_valid_patchnet}%
\end{table}%


{\bf RQ2: How effective is PatchNet compared to other state-of-the-art stable patch identification models?}  
\textcolor{black}{To answer this RQ, we use the five test sets of the dataset described in RQ1. Of these, we take one test set as the testing data and regard the remaining patches as the training data. We repeat this five times, and then average the results to get the aggregated accuracy, precision, recall, F1, and AUC scores.}
Table~\ref{tab:msg_code} shows the results for PatchNet and the other
baselines. PatchNet achieves average accuracy, precision, recall, F1 score, and AUC of 0.862, 0.839, 0.907, 0.871, and 0.860, respectively. Compared to the best
performing baseline, F-NN, these constitute improvements of
6.55\%, 0.12\%, 16.13\%, 7.80\%, and 6.30\%, respectively. PatchNet thus achieves
about the same precision as F-NN, but a significant improvement in terms of
recall. This is achieved without the feature engineering required for
the F-NN approach, but rather by automatically learning the weight of the filters
via our hierarchical deep learning-based architecture. 

We also employ Scott-Knott ESD ranking~\cite{tantithamthavorn2017empirical} to
statistically compare the performance of PatchNet and the four considered
approaches (i.e., PatchNet, F-NN, LS-CNN, and LPU-SVM). The results show
that PatchNet consistently appears in the top Scott-Knott ESD rank in terms
of accuracy, precision, recall, F1 score, and AUC. The ranks of the four
considered approaches are furthermore consistent (i.e., PatchNet $>$ F-NN $>$ LS-CNN $>$ LPU-SVM) except for recall (i.e., PatchNet $>$ LS-CNN $>$ F-NN $>$ LPU-SVM).



\begin{table}[t!]
  \centering
  \caption{PatchNet vs. Keyword, LPU+SVM, LS-CNN, and F-NN.}
    \begin{tabular}{|l|c|c|c|c|c|}
    \hline
          & \textbf{Accuracy} & \textbf{Precision } & \textbf{Recall} & \textbf{F1} & \textbf{AUC} \\
    \hline
    \hline
    Keyword & 0.626 & 0.683 & 0.515 & 0.587 & 0.630 \\
    \hline
    LPU+SVM & 0.731 & 0.751 & 0.716 & 0.733 & 0.731 \\
    \hline
    LS-CNN & 0.765 & 0.766 & 0.785 & 0.775 & 0.765 \\
    \hline
    F-NN & 0.809 & 0.838 & 0.781 & 0.808 & 0.809 \\
    \hline
    PatchNet & \textbf{0.862} & \textbf{0.839} & \textbf{0.907} & \textbf{0.871} & \textbf{0.860} \\
    \hline
    \end{tabular}%
  \label{tab:msg_code}%
\end{table}%



\begin{figure}
\centering
\includegraphics[scale=0.375]{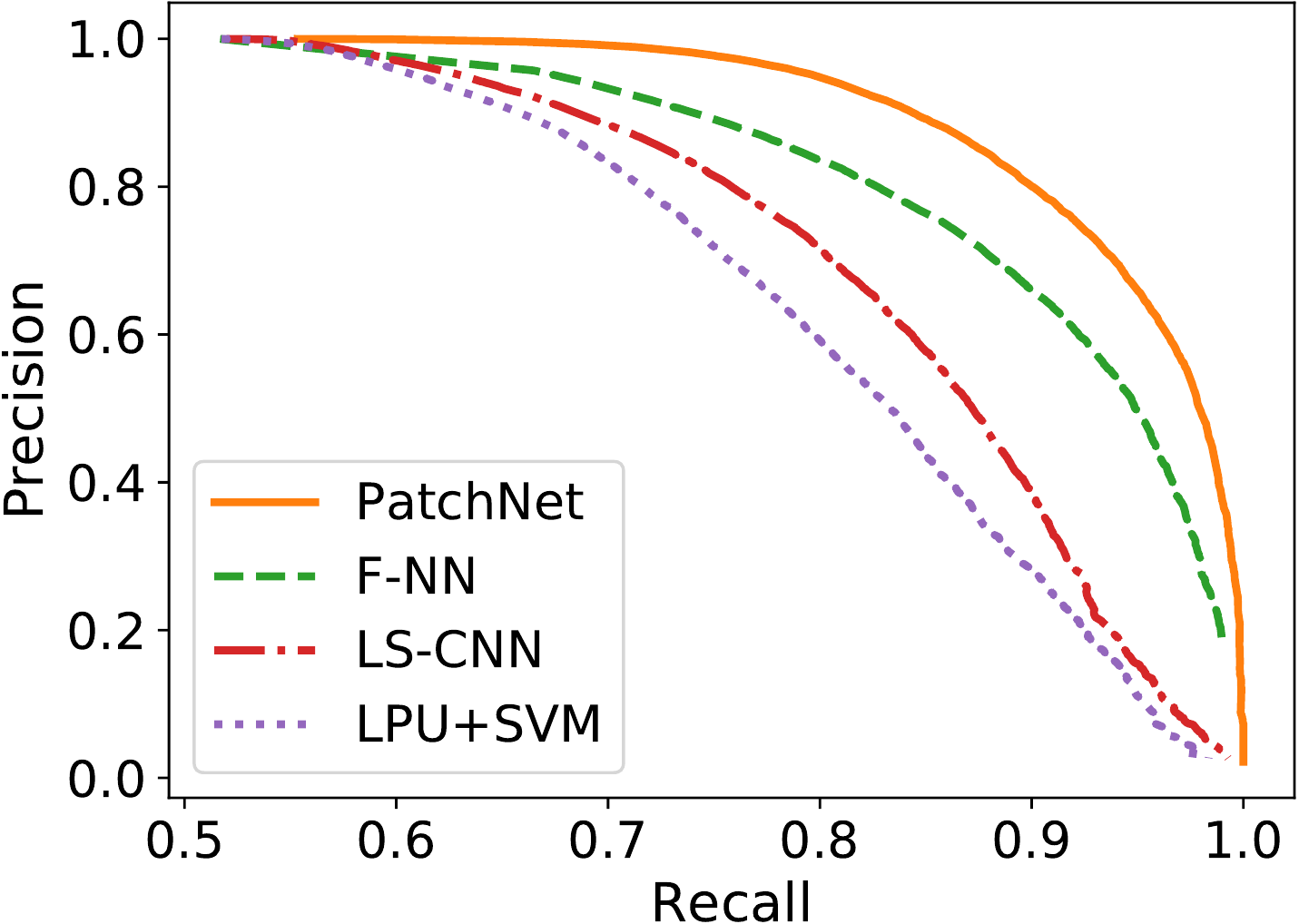}
	\caption{Precision-recall curve: PatchNet vs. LPU+SVM, LS-CNN, and F-NN.}
	\label{fig:prc_rc_curve}
\end{figure}

Fig.~\ref{fig:prc_rc_curve} compares the precision-recall curves for
PatchNet and the baselines. For most values
on the curve, PatchNet obtains the highest recall for a given precision and the highest precision for a given recall. For example, for a low false positive rate of 5 percent (precision of 0.95), PatchNet achieves a recall of 0.786 which is 14.9\% higher than that of the best performing baseline. Likewise, for a low false negative rate of 5 percent (recall of 0.95), PatchNet achieves a precision of 0.603 which is 41.2\% higher than that of the best performing baseline. In addition, considering the sweet spots where both precision and recall are high (larger than 0.8), PatchNet can achieve an F1 score
of up to 0.886 which is 10.6\% higher than that of the best performing baseline.



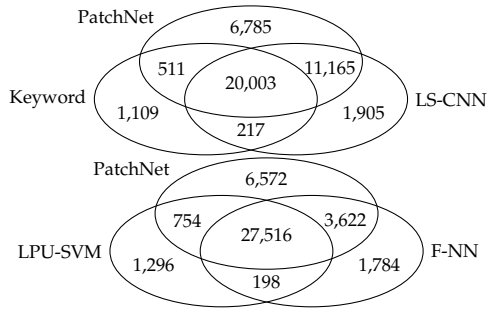
\begin{figure}
\def\firsta{(-.2,0) ellipse (6em and 3em)}
\def\seconda{(1,0) ellipse (6em and 3em)}
\def\thirda{(0.4,0.5) ellipse (6em and 3em)}
\centerline{\begin{tikzpicture} \scriptsize
    \draw \firsta node [below] (An) {};
    \draw \seconda node [below] (Bn) {};
    \draw \thirda node [above] (Cn) {};
%
    \node at (-1.1,-.2) (A) {1,109};
    \node at (1.9,-.2) (B) {1,905};
    \node at (.4,.95) (C) {6,785};
    \node at (.4,.2) {20,003};
    \node at (.4,-.4) {217}; 
    \node at (-0.65,.4) {511}; 
    \node at (1.45,.4) {11,165}; 
    \node[anchor=east] at (-1.7,0) {Keyword};
    \node[anchor=west] at (2.5,0) (B) {LS-CNN};
    \node[anchor=east] at (-.8,1.1) (C) {PatchNet};
  \end{tikzpicture}}


\def\first{(-.2,0) ellipse (6em and 3em)}
\def\second{(1,0) ellipse (6em and 3em)}
\def\third{(0.4,0.5) ellipse (6em and 3em)}
\centerline{\begin{tikzpicture} \scriptsize
    \draw \first node [below] (An) {};
    \draw \second node [below] (Bn) {};
    \draw \third node [above] (Cn) {};
%
    \node at (-1.1,-.2) (A) {1,296};
    \node at (1.9,-.2) (B) {1,784};
    \node at (.4,.95) (C) {6,572};
    \node at (.4,.2) {27,516};
    \node at (.4,-.4) {198}; 
    \node at (-0.65,.4) {754}; 
    \node at (1.45,.4) {3,622}; 
    \node[anchor=east] at (-1.7,0) {LPU-SVM};
    \node[anchor=west] at (2.5,0) (B) {F-NN};
    \node[anchor=east] at (-.8,1.1) (C) {PatchNet};
  \end{tikzpicture}}
	\caption{Venn diagrams showing the number of stable
          patches identified by PatchNet and the various baselines}
	\label{fig:venn_diagram}
\end{figure}

Fig.~\ref{fig:venn_diagram} shows Venn diagrams indicating the number of patches that PatchNet and each of the baselines correctly recognize as stable. The top diagram compares the Keyword approach to the two approaches, PatchNet and LS-CNN, that automatically learn the relevant features.  While there are over 20K patches that all three approaches classify as stable, there are another 11K that are
found by both learning-based approaches, showing the advantage of
learning-based approach.  
As compared to Keyword and LS-CNN, there are almost 7,000 patches that are only recognized by PatchNet, while this is the case for fewer than 2,000 patches for LS-CNN, showing the value of an approach
that takes the properties of code changes into account.

The bottom diagram
then compares PatchNet to the two approaches, LPU+SVM and F-NN, in which
the code features are handcrafted. 
While all three approaches correctly
recognize over 27K patches as stable, there are again 3x more
patches that only PatchNet correctly detects as stable than there
are that only each of the other two approaches recognizes as stable. Examples of PatchNet true
positives not found by the other baselines include 5567e989198b\footnote{git://git.kernel.org/pub/scm/linux/kernel/git/torvalds/linux.git} and 2e31b4cb895a.
Examples of PatchNet false negatives found by at least one other baseline include
03f219041fdb and 56199016e867.

\textcolor{black}{All of the above measures of precision and recall assume that the set of
patches found in the tested Linux kernel versions is the correct one.  Our
motivation, however, is that bug fixing patches that should be propagated
to the Linux kernel stable versions are being overlooked by the existing
manual labeling process.  Showing that PatchNet improves on the existing
manual process requires collecting a dataset of patches that have not been
propagated to stable kernels, but should have been.  Collecting such a
dataset, however, requires substantial Linux kernel expertise, which it is
not feasible to harness at a large scale.  We have nevertheless been able to
carry out two experiments in this direction.  First, we randomly selected
200 patches predicted as stable patches by PatchNet, but that were not
marked as stable in our dataset.  We sent the 200 patches to Sasha Levin (a
Linux stable-kernel maintainer and the developer of F-NN) to label. Among
the 200 patches, Levin labeled 61 patches (i.e., 30.5\%) as stable,
highlighting that our approach can find many additional stable patches that
were not identified by the existing manual process.  Note that these
patches predated Levin's used of F-NN on the Linux kernel.  Second, we
looked at commits that have no Cc stable tag that Sasha Levin selected with
the aid of F-NN for the Linux 4.14 stable tree.  These commits postdate all
of the commits in our dataset.  There are over 1,800 of them, showing the
false negatives in the existing manual process and the need for automated
support.  PatchNet detects 91\% of them as stable.  The relationship
between the results of F-NN and PatchNet is similar to that shown in
Fig.~\ref{fig:venn_diagram} for patches in our original dataset and
confirms that PatchNet can find stable patches that were not identified by
the existing manual process.}

{\bf RQ3: Does PatchNet benefit from considering
  both the commit message and the code changes, and do function names help
  identify stable patches?}
  To answer this RQ, we conduct an ablation
test~\cite{korbar2017deep, liu2017deep}
by ignoring the
commit message, the code changes, or the function names in the code changes
in a given patch one-at-a-time and evaluating the performance. We create
three variants of PatchNet: PatchNet-C, PatchNet-M, and
PatchNet-NN. PatchNet-C uses only code change information while PatchNet-M
uses only commit message information. PatchNet-NN uses both code change and
commit message information, but ignores the function names in the code
changes. We again use the five copies of the dataset described in RQ1 and
compute the various evaluation metrics.

Table~\ref{tab:components} shows that the performance of PatchNet degrades
if we ignore any one of the considered types of information. Accuracy,
precision, recall, F1 score, and AUC drop by 19.39\%, 15.41\%, 21.26\%, 18.34\%,
and 16.06\% respectively if we ignore commit messages. They drop by 16.96\%,
14.62\%, 16.58\%, 14.76\%, and 14.21\% respectively if we ignore code
changes.  And they drop by 11.08\%, 12.62\%, 16.43\%, 13.86\%, and 11.98\%
respectively if we ignore function names.  Thus, each kind of information
contributes to PatchNet's performance.  Additionally, the drops are
greatest if we ignore commit messages, indicating that they are slightly
more important than the other two to PatchNet's performance.

\begin{table}[t!]
  \centering
  \caption{Contribution of commit messages,  code changes and function names to PatchNet's performance}
    \begin{tabular}{|l|c|c|c|c|c|}
    \hline
          & \textbf{Accuracy} & \textbf{Precision} & \textbf{Recall} & \textbf{F1}    & \textbf{AUC} \\
    \hline
    \hline
    PatchNet-C & 0.722 & 0.727 & 0.748 & 0.736 & 0.741 \\
    \hline
    PatchNet-M & 0.737 & 0.732 & 0.778 & 0.759 & 0.753 \\
    \hline
    PatchNet-NN & 0.776 & 0.745 & 0.779 & 0.765 & 0.768 \\
    \hline
    PatchNet & \textbf{0.862} & \textbf{0.839} & \textbf{0.907} & \textbf{0.871} & \textbf{0.860} \\
    \hline
    \end{tabular}%
  \label{tab:components}%
\end{table}%

\textcolor{black}{{\bf RQ4: What are the results of PatchNet on the complete
    set of Linux kernel patches?} For RQ1, we use a dataset collected such
  that the number of stable and non-stable patches is roughly
  balanced. Among the 267,251 patches that meet the selection criteria, we
  picked 42,408 stable patches and 39,995 non-stable patches to build our
  dataset. To investigate the results of PatchNet on the complete set of
  patches from Linux v3.0-v4.12 having at most 100 lines (and accepted by
  our preprocessor), we randomly divide the remaining 184,481 non-stable
  patches into five sets and merge each of them with each of the five test
  sets described in RQ1. After this process, we have a new collection of
  five test sets. In each test set, there are around 8.4K stable patches
  and 44.8K non-stable patches. For each new test set, we use the
  corresponding model trained for RQ1. We repeat this five times, and then
  average the results to get the aggregated AUC score. PatchNet achieves an
  average AUC of 0.808. Since the new five test sets are highly imbalanced
  (only 15.79\% patches are stable patches), we omit the other metrics
  (i.e., accuracy, precision, recall, and F1)~\cite{nguyen2009learning, tantithamthavorn2018impact, mcintosh2017fix}.
  We also trained PatchNet on a whole training dataset (i.e., 42,408 stable
  patches and 39,995 non-stable patches) and evaluated it on 184,481
  non-stable patches. We find that PatchNet can correctly label them as
  non-stable 81.32\% of the time.}

\textcolor{black}{We also check the effectiveness of PatchNet on patches that
  have more than 100 lines of code (i.e., long patches). As mentioned
  earlier, we omit those patches from our training dataset as they do not
  meet the selection criteria of Linux kernel. We collect 52,415 long
  patches from July 2011 to July 2017. Among them, there are 3,376 long
  stable patches and 49,039 long non-stable patches. 21.33\% of these
  patches contain the ``Cc: stable'' tag.  The others may have been
  manually selected for stable versions despite not having a tag or may
  come from the release candidates. We again train PatchNet on the whole
  training dataset and evaluate the effectiveness of PatchNet on the 52,415
  long patches. PatchNet achieves an AUC score of 0.805. Again we only use
  AUC as this dataset is highly imbalanced~\cite{nguyen2009learning,
    tantithamthavorn2018impact, mcintosh2017fix}.}

\textcolor{black}{Finally, we also check whether there is a difference of
  performance in classifying patches containing a ``Cc: stable'' tag and
  patches that do not containing a ``Cc: stable'' tag. Among the 42,408
  stable patches, there are 15,410 stable patches with a stable tag and
  26,998 stable patches with no stable tag. The latter may again have been
  manually selected for stable versions despite not having a tag or may
  come from the release candidates.  For each test set described in RQ1, we
  split the stable patches into two groups: tagged stable patches and
  non-tagged stable patches. We run PatchNet on the stable patches of each
  test set to predict the stable patches and sum the results of predicting
  the stable patches.  Among 15,410 tagged stable patches, PatchNet
  predicts 14,578 patches as stable patches (i.e., 94.60\%).  Among the
  26,998 non-tagged stable patches, PatchNet predicts 23,466 patches as
  stable patches (i.e., 86.92\%). We find that PatchNet is more successful
  at recognizing tagged patches, even when it does not have access to
  information about the ``Cc: stable'' tag.}

\section{Qualitative Analysis and Discussion}
\label{sec:qualitative_discussion}

  
  \textcolor{black}{In this section, we analyze some of the
    results obtained in Section~\ref{sec:rq}, considering in detail a patch
    where PatchNet performs well and another where it performs poorly.}

\subsection{Successful Case}
\label{sec:good_case}

\begin{figure}[t!]
\begin{lstlisting}[language=diff]
commit 203dc2201326fa64411158c84ab0745546300310
Author: Jakob Bornecrantz <jakob@vmware.com>
Date:   Mon Sep 17 00:00:00 2001 +0000

    vmwgfx: Do better culling of presents
    
    Signed-off-by: Jakob Bornecrantz <jakob@vmware.com>
    Reviewed-by: Thomas Hellstrom <thellstrom@vmware.com>
    Signed-off-by: Dave Airlie <airlied@redhat.com>

diff --git a/drivers/gpu/drm/vmwgfx/vmwgfx_kms.c 
           b/drivers/gpu/drm/vmwgfx/vmwgfx_kms.c
index ac24cfd..d31ae33 100644
--- a/drivers/gpu/drm/vmwgfx/vmwgfx_kms.c
+++ b/drivers/gpu/drm/vmwgfx/vmwgfx_kms.c
@@ -1098,6 +1098,7 @@ int vmw_kms_present(struct vmw_private *dev_priv,
            ...
+	int left, right, top, bottom;
            ...
+	left = clips->x;
+	right = clips->x + clips->w;
+	top = clips->y;
+	bottom = clips->y + clips->h;
+
+	for (i = 1; i < num_clips; i++) {
+		left = min_t(int, left, (int)clips[i].x);
+		right = max_t(int, right, (int)clips[i].x + clips[i].w);
+		top = min_t(int, top, (int)clips[i].y);
+		bottom = max_t(int, bottom, (int)clips[i].y + clips[i].h);
+	}
+                       return err;
            ...
-	cmd->body.srcRect.left = 0;
-	cmd->body.srcRect.right = surface->sizes[0].width;
-	cmd->body.srcRect.top = 0;
-	cmd->body.srcRect.bottom = surface->sizes[0].height;
+	cmd->body.srcRect.left = left;
+	cmd->body.srcRect.right = right;
+	cmd->body.srcRect.top = top;
+	cmd->body.srcRect.bottom = bottom;
            ...
-		blits[i].left   = clips[i].x;
-		blits[i].right  = clips[i].x + clips[i].w;
-		blits[i].top    = clips[i].y;
-		blits[i].bottom = clips[i].y + clips[i].h;
+		blits[i].left   = clips[i].x - left;
+		blits[i].right  = clips[i].x + clips[i].w - left;
+		blits[i].top    = clips[i].y - top;
+		blits[i].bottom = clips[i].y + clips[i].h - top;
            ...
-		int clip_x1 = destX - unit->crtc.x;
-		int clip_y1 = destY - unit->crtc.y;
-		int clip_x2 = clip_x1 + surface->sizes[0].width;
-		int clip_y2 = clip_y1 + surface->sizes[0].height;
+		int clip_x1 = left + destX - unit->crtc.x;
+		int clip_y1 = top + destY - unit->crtc.y;
+		int clip_x2 = right + destX - unit->crtc.x;
+		int clip_y2 = bottom + destY - unit->crtc.y;
            ...
\end{lstlisting}
\caption{Example of a successfully identified stable patch.}
\label{fig:good_case}
\end{figure}


\textcolor{black}{We first present a patch that PatchNet can predict as a stable patch, intending to show an advantage of our model.}

\textcolor{black}{Figure~\ref{fig:good_case} shows a patch propagated to stable
  kernels. The commit message is on line 5 and the code changes are on
  lines 16-59. The code changes include one changed file, five hunks, 12
  removed lines, and 25 added lines. PatchNet is able to predict the patch
  in Figure~\ref{fig:good_case} as stable patch. We see that the commit
  message of this patch is quite short and does not contain keywords such
  as ``bug'' or ``fix''. To recognize the patch as a stable patch, the
  stable kernel maintainer has to study the code changes to understand the
  impact of the changes in the kernel code. In the code changes, the four
  variables (i.e, {\tt left}, {\tt right}, {\tt top}, and {\tt bottom}) are
  defined and used across the multiple hunks in the changed file (i.e.,
  {\tt vmwgfx\_kms.c}). We also see the difference between removed lines
  and added lines when the author committed his code. By representing the
  removed code and the added code as two three-dimensional matrices (each
  dimension represents the number of hunks, the number of removed or added
  code lines, and the number of words in each removed or added code line),
  PatchNet uses the \textit{removed code module} and the \textit{added code
    module} to construct the embedding vector of the removed code and added
  code, respectively (see Section~\ref{sec:commit_code_model}). The two
  embedding vectors are then concatenated to represent the code change
  information. By doing this process, the distinction between
    removed lines and
  added lines is preserved. PatchNet automatically learns from this rich
  representation by updating its parameters during the training process
  (see Section~\ref{sec:parameters}) to build a model that can predict
  whether a patch is stable.}


\textcolor{black}{On the other hand, we find that none of the other baselines are able to classify the patch in Figure~\ref{fig:good_case} as a stable patch. \textit{Keyword} is a heuristic approach that only looks at whether the content of a commit message includes ``bug'' or ``fix''. \textit{LS-CNN} concatenates the removed lines and added lines in the multiple hunks without preserving the code changes information. \textit{LPU+SVM} and \textit{F-NN} define a set of features for the code changes (i.e., the number of removed code lines, the number of added code lines, the number of hunks in a commit, etc.). The manual creation of code changes features may overlook features that are important to identify stable patches, making \textit{LPU+SVM} and \textit{F-NN} unable to classify the patch in Figure~\ref{fig:good_case} as a stable patch. }

\subsection{Unsuccessful Case}
\label{sec:bad_case}

\begin{figure}[t!]
\begin{lstlisting}[language=diff]
commit 	c607f450f6e49f5794f27617bedc638b51044d2e 
Author: Al Viro <viro@zeniv.linux.org.uk>
Date:   Sat May 11 12:38:38 2013 -0400 

    au1100fb: VM_IO is set by io_remap_pfn_range()
    
    Signed-off-by: Al Viro <viro@zeniv.linux.org.uk>

diff --git a/drivers/video/au1100fb.c b/drivers/video/au1100fb.c
index 700cac067b46..ebeb9715f061 100644
--- a/drivers/video/au1100fb.c
+++ b/drivers/video/au1100fb.c
@@ -385,8 +385,6 @@ int au1100fb_fb_mmap(struct fb_info *fbi, struct vm_area_struct *vma)
 	vma->vm_page_prot = pgprot_noncached(vma->vm_page_prot);
 	pgprot_val(vma->vm_page_prot) |= (6 << 9); //CCA=6

-	vma->vm_flags |= VM_IO;
-
 	if (io_remap_pfn_range(vma, vma->vm_start, off >> PAGE_SHIFT,
 				vma->vm_end - vma->vm_start,
 				vma->vm_page_prot)) {
\end{lstlisting}
\caption{Example of an unsuccessfully identified stable patch.}
\label{fig:bad_case}
\end{figure}

\textcolor{black}{Next, we present a patch that PatchNet fails to classify
  correctly as a stable patch. This example serves to provide an
  understanding of cases in which PatchNet may not perform well.}

\textcolor{black}{Figure~\ref{fig:bad_case} shows a stable patch that was not recognized
  by PatchNet. Its commit message does not contain any keywords (i.e.,
  ``bug'' or ``fix'') that suggest whether the patch is a stable patch. The
  code changes only include one removed line and the removed line contains
  only three words: {\tt vma}, {\tt vm\_flags}, and {\tt VM\_IO}. As there
  is very little information in both the commit message and the code
  changes, PatchNet is unable to predict the patch in
  Figure~\ref{fig:bad_case} as a stable patch. We find that the other baselines
  (i.e, \textit{keywords}, \textit{LS-CNN}, and \textit{LPU+SVM}), except
  \textit{F-NN}, also fail to classify the patch as a stable
  patch. \textit{F-NN} considers not only the commit message and the code
  changes of the given patch, but also information such as author name,
  reviewer information, file names, etc. 
  This suggests that when the information of the commit message
  and the code changes is limited, an approach that takes advantage of
  other information in a given patch may perform better than PatchNet.
  }
 
\section{Threats to Validity}
\label{sec:threat}

\noindent{\bf Internal validity}. Threats to internal validity relate to
errors in our experiments and experimenter bias. We have double checked our
code and data, but errors may remain.
In
the baseline approach by Tian et al.~\cite{tian2012identifying}, commits
were labeled by an author with expertise in Linux kernel code, which may
introduce author bias. In this work, none of the authors label the commits.

\noindent {\bf External validity}. Threats to external
validity relate to the generalizability of our approach. We have evaluated
our approach on more than 80,000 patches. We believe this is a good number
of patches. Still, the results may differ if we consider other sets of
Linux kernel patches. Similar to the evaluation of Tian et
al.~\cite{tian2012identifying}, we only investigated Linux kernel patches,
although PatchNet can be applied to patches of other systems, if
labels are available. In the
future, we would like to consider more projects.
Still, we note that the Linux kernel represents one of the largest open
source projects, with over 16 million lines of C code, and that different
kernel subsystems have different developers and very different purposes, resulting in a wide variety
of code.

\noindent {\bf Construct validity}. Threats to construct validity relate to the suitability of our evaluation metrics. We use standard metrics commonly used to evaluate classifier performance. Thus, we believe there is little threat to construct validity.

\section{Related Work}
\label{sec:related_work}

Researchers have applied deep learning techniques to solve software
engineering problems, including code clone
detection~\cite{white2016deep,li2017cclearner,bui2018hierarchical}, software traceability link
recovery~\cite{guo2017semantically}, bug
localization~\cite{huo2016learning, lam2017bug}, defect
detection~\cite{yang2015deep, wang2016automatically}, automated program
repair~\cite{gupta2017deepfix}, and API
learning~\cite{gu2016deep}. However, we did not find any research that applied
deep learning techniques to learn semantic representations of patches for
similar tasks such as stable patch identification, patch classification, etc. Here, we briefly describe the most closely related work besides the baselines described in Section~\ref{sec:baselines}.

\noindent {\bf Sequence-to-sequence learning.} Gu \emph{et al.} adopted a neural language model named a Recursive Neural Network (RNN)~\cite{hagan1996neural, mikolov2010recurrent} encoder-decoder to generate API usage sequences, \emph{i.e.,} a sequence of method names, for a given natural language query~\cite{gu2016deep}. Gupta \emph{et al.} proposed DeepFix to automatically fix syntax errors in C code~\cite{gupta2017deepfix}. DeepFix leverages a multi-layered sequence-to-sequence neural network with attention~\cite{mnih2014recurrent}, to process the input code and a decoder RNN with attention that generates the output fixed code. The above studies focus on learning sequence-to-sequence mappings and thus consider a different task than the one considered in our work.

\noindent {\bf Learning code representation.}
CCLearner~\cite{li2017cclearner} learns a deep neural network classifier
from clone pairs and non clone pairs to detect clones.  To represent code,
it extracts features based on different categories (reserved words,
operators, etc.) of tokens in source code. White \emph{et al.} presented another
deep learning-based clone detector~\cite{white2016deep}. Their tool first
uses RNN to map program tokens to continuous-valued vectors, and then uses
RNN to combine the vectors with extracted syntactic features to train a
classifier. Wang \emph{et al.} used a deep belief network
(DBN)~\cite{hinton2009deep} to predict defective
code~\cite{wang2016automatically}. The DBN learns a semantic representation
(in the form of a continuous-valued vector) of each source code file from
token vectors extracted from programs' ASTs. Lam \emph{et al.} combined deep
learning with information
retrieval to localize
buggy files based on bug reports~\cite{lam2017bug}. Bui and Jiang
proposed a deep learning based approach to automatically learn
cross-language representations for various kinds of structural code
elements (\emph{i.e.,} expressions, statements, and methods) for program
translation~\cite{bui2018hierarchical}. Different from the above
studies, we design a novel deep learning architecture that focuses on code
changes, taking into account their hierarchical and structural properties.


\noindent {\bf Learning of both code and text representations.} Huo and Li proposed a model, LS-CNN, for classifying if a source code file is related to a bug report (\emph{i.e.,} the source code file needs to be fixed to resolve the bug report)~\cite{huo2017enhancing}. LS-CNN is the first code representation learning method that combines CNN and LSTM (a specific type of RNN) to extract semantic representations of both code (in their case: a source code file) and text (in their case: a bug report). Similar to LS-CNN, PatchNet also learns semantic representations of both code and text. However, different from LS-CNN, PatchNet includes a new representation learning architecture for commit code comprising the representations of removed code and added code of each affected file in a given patch. The representation of removed code and added code is able to capture the sequential nature of the source code inside a code change, and it is learned following a CNN-3D architecture~\cite{ji20133d} instead of LSTM. Our results in Section~\ref{sec:exp} show that PatchNet can achieve an 11.24\% improvement in terms of F1 over the LS-CNN model.


\vspace{-0.1cm}
\section{Conclusion}
\label{sec:conclusion}

In this paper, we propose PatchNet, a hierarchical deep learning-based model for identifying stable patches in the Linux kernel. For each patch, our model constructs embedding vectors from the commit message and the set of code changes. The embedding vectors are concatenated and then used to compute a prediction score for the patch. Different from existing deep learning techniques working on the source code~\cite{white2016deep, wang2016automatically, huo2017enhancing, li2017cclearner, guo2017semantically, lam2017bug, gu2016deep}, our hierarchical deep learning-based architecture takes into account the structure of code changes (\emph{i.e.,} files, hunks, lines) and the sequential nature of source code (by considering each line of code as a sequence of words) to predict stable patches in the Linux kernel. 


We have extensively evaluated PatchNet on a new dataset containing 82,403 recent Linux kernel patches. On this dataset, PatchNet outperforms 
four baselines including two also based on deep-learning. In particular, for a wide range of values in the precision-recall curve, PatchNet obtains the highest recall for a given precision, as well as the highest precision for a given recall. For example, PatchNet achieves a 14.9\% higher recall (0.786) at a high precision level (0.950) and a 41.2\% higher precision (0.603) at a high recall level (0.950) compared to the best-performing baseline.


In future work, we want to investigate ways to improve
our approach further, {\em e.g.}, by incorporating additional data such as
more kinds of names and type information.  Another issue is to
identify the stable versions to which a patch should be applied. We plan to
investigate whether machine learning can help with this issue.  It would
also be interesting to apply our approach that learns patch embeddings
to other related problems, {\em e.g.} identification of valid/invalid patches in
automated program repair~\cite{xiong2018identifying}, assignment of patches
to developers for code review~\cite{thongtanunam2015should,
zanjani2016automatically}, etc.

\noindent \textbf{Dataset and Code.} The dataset and code for PatchNet are available at \url{https://github.com/hvdthong/PatchNetTool}. A video demonstration of PatchNet is available at \url{https://goo.gl/CZjG6X}.

\noindent{\bf Acknowledgement.} This research was supported by the Singapore National Research Foundation (Award number: NRF2016-NRF-ANR003) and the ANR ITrans project.

\balance

\let\oldthebibliography=\thebibliography
\let\endoldthebibliography=\endthebibliography
\renewenvironment{thebibliography}[1]{%
  \vspace{0pt}
  \begin{oldthebibliography}{#1}%
    \setlength{\parskip}{0ex}%
    \setlength{\itemsep}{0.5ex}%
}%
{%
  \end{oldthebibliography}%
}
{\scriptsize
\def\baselinestretch{0.95}
\bibliographystyle{IEEEtranS}

\bibliography{references}
}

\begin{figure*}	[t!]
	\begin{subfigure}{0.475\textwidth}
		\begin{minipage}{1.0\columnwidth}			
			\begin{tabular}{p{1.0\columnwidth}}
				\begin{IEEEbiography}[{\includegraphics[width=1in,height=1.25in,clip]{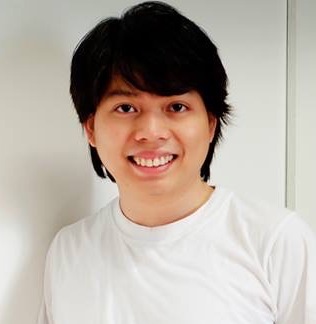}}]%
					{Thong Hoang} is a Ph.D. candidate at the School of Information  Systems, Singapore Management University (SMU), advised by Associate Professor David Lo. He received his B.Eng. and M.S. in Computer Science from the Ho Chi Minh University of Technology - Vietnam and Konkuk University - South Korea, respectively. He worked as a research engineer at Ghent University and the Living Analytics Research Centre, SMU. His research interests include software engineering and machine learning, with particular focus on software bug localization, defect prediction, and program representation. 
				\end{IEEEbiography}
			\end{tabular}
		\end{minipage} 
	\end{subfigure} \hfill
	\begin{subfigure}{0.475\textwidth}
		\begin{minipage}{1.0\columnwidth}			
			\begin{tabular}{p{1.0\columnwidth}}
				\begin{IEEEbiography}[{\includegraphics[width=1in,height=1.25in,clip]{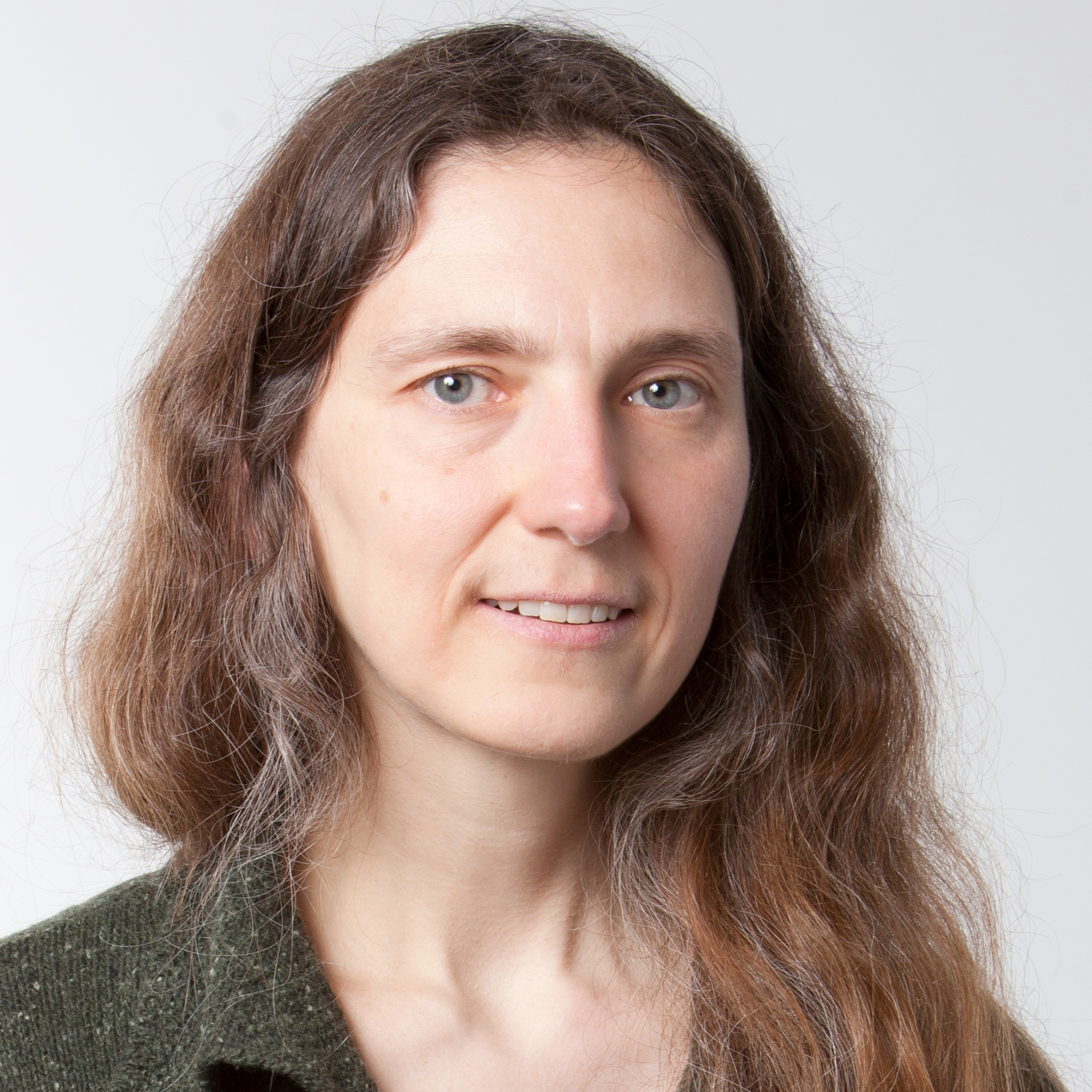}}]%
					{Julia Lawall} received the Ph.D. degree in 1994 from the Indiana University. She has been a senior research scientist at INRIA since 2011. Previously, she was an Associate Professor at the University of Copenhagen.  Her research interests are in the area of improving the quality of infrastructure software, using a variety of approaches including program analysis, program transformation, and the design of domain-specific languages.  She develops the tool Coccinelle that is extensively used in the development of the Linux kernel and she has over 2000 patches in the Linux kernel based on this work.
				\end{IEEEbiography}
			\end{tabular}
		\end{minipage}
	\end{subfigure} 
\end{figure*}

\begin{figure*} [t!]
	\begin{subfigure}{0.475\textwidth}
		\begin{minipage}{1.0\columnwidth}			
			\begin{tabular}{p{1.0\columnwidth}}
				\begin{IEEEbiography}[{\includegraphics[width=1in,height=1.25in,clip]{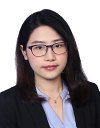}}]%
					{Yuan Tian} is an Assistant Professor at the School of Computing, Queen's University, Canada. She received her Ph.D. in Information Systems from Singapore Management University (SMU) in 2017, with extended visits to Carnegie Mellon University and INRIA (French National Institute for Research in Computer Science and Control) Paris. Prior to joining Queen's, Dr. Tian worked at the Living Analytics Research Centre (LARC) of SMU as a Data Scientist. Her research interests include software engineering and data mining, with particular focus on mining software repositories and automated software engineering. She currently holds an NSERC Research Grant (2019-2024) entitled Reliable and Explainable Recommender Systems for Efficient Software Development. Dr. Tian has published over 30 papers in software engineering journals and conference proceedings, including ICSE, ASE, TOSEM, and EMSE. More information can be found at \url{http://sophiaytian.com}.
				\end{IEEEbiography}
			\end{tabular}
		\end{minipage} 
	\end{subfigure} \hfill
	\begin{subfigure} [t!] {0.475\textwidth}
		\begin{minipage}[t!]{1.0\columnwidth}			
			\begin{tabular} [t!] {p{1.0\columnwidth}}
				\begin{IEEEbiography}[{\includegraphics[width=1in,height=1.25in,clip]{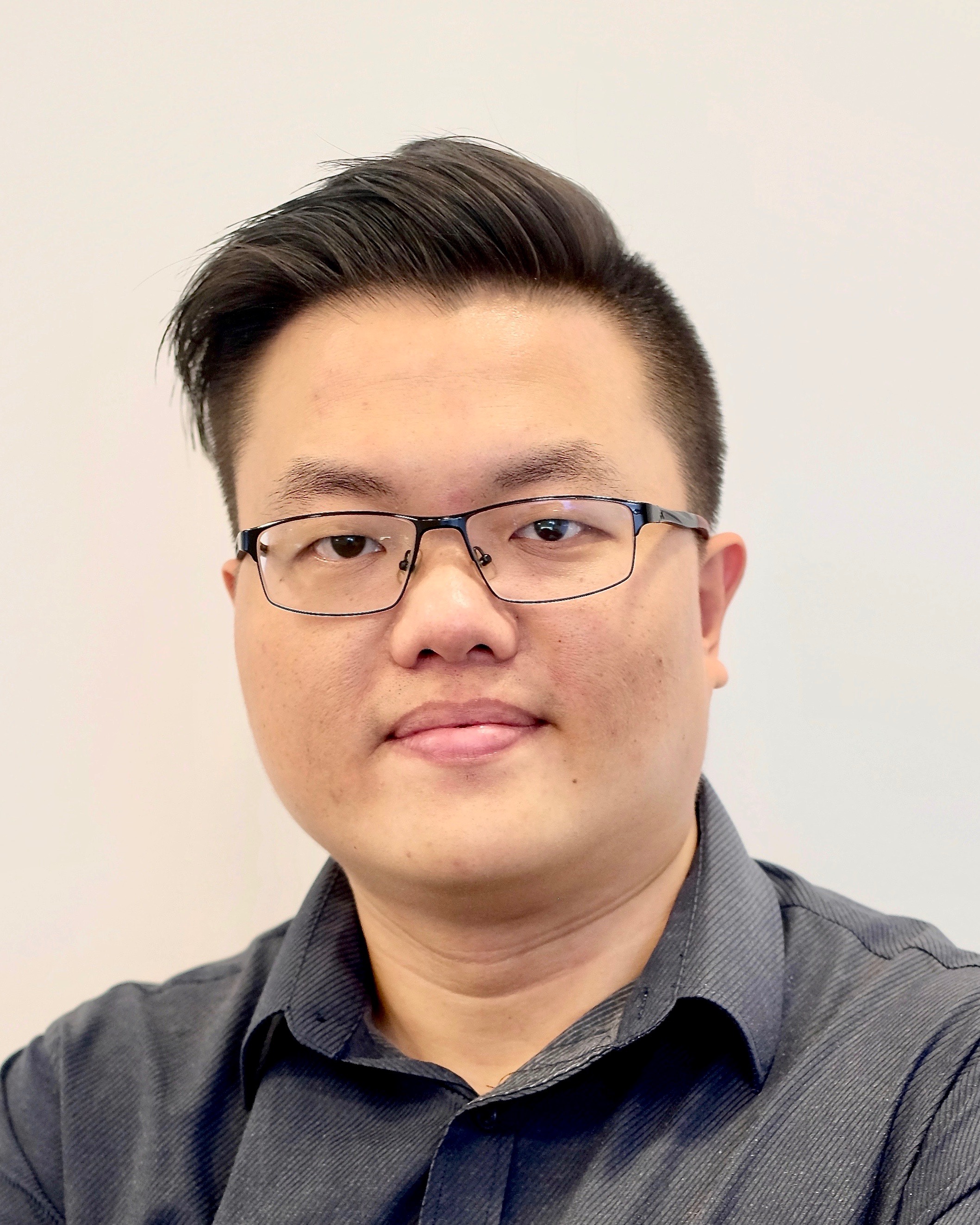}}]%
					{Richard J. Oentaryo} is a Senior Data Scientist at McLaren Applied Technologies, currently serving as Analytics Lead and Technical Pre-sales Support for Condition Insight projects in Asia Pacific. Previously, he was a Research Scientist at the Living Analytics Research Centre (LARC), Singapore Management University (SMU) in 2011-2016, and Research Fellow at the School of Electrical and Electronic Engineering, Nanyang Technological University (NTU) in 2010-2011. He received Ph.D. and B.Eng. (First Class Honors) from the School of Computer Engineering (SCE), NTU, in 2011 and 2004 respectively. Dr. Oentaryo has published 40 papers in international journal and conference venues, and his research interests span machine learning, information retrieval, optimisation, and nature-inspired computing. Further information is available at \url{http://sites.google.com/site/oentaryo/}.  
					\bigskip
					\bigskip
				\end{IEEEbiography}
			\end{tabular}
		\end{minipage}
	\end{subfigure} 
\end{figure*}

\begin{figure*}
	\begin{subfigure}{0.475\textwidth}
		\begin{minipage}[!t]{1.0\columnwidth}			
			\begin{tabular}{p{1.0\columnwidth}}
				\begin{IEEEbiography}[{\includegraphics[width=1in,height=1.25in,clip]{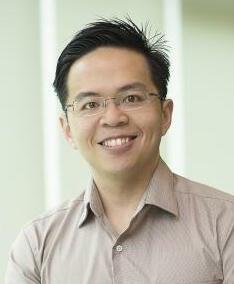}}]%
					{David Lo} obtained his PhD degree from the School of Computing, National University of Singapore in 2008. He is currently an Associate Professor in the School of Information Systems, Singapore Management University. He has more than 10 years of experience in software engineering and data mining research and has more than 100 publications in these areas. He received the Lee Foundation, Lee Kong Chian, and Lee Kuan Yew Fellowships from the Singapore Management University in 2009, 2018, and 2019 respectively. He has received a number of international research awards including five ACM SIGSOFT Distinguished Paper awards for his work on software analytics. He has served as general and program co-chair of several well-known international conferences (e.g., IEEE/ACM International Conference on Automated Software Engineering), and editorial board member of a number of high-quality journals (e.g., Empirical Software Engineering). He is a Distinguished Member of ACM and Senior Member of IEEE.
				\end{IEEEbiography}
			\end{tabular}
		\end{minipage} \hfill
	\end{subfigure} 
\end{figure*}

\end{document}